\documentclass[pra,twocolumn,amsmath,amssymb,superscriptaddress]{revtex4-1}

\usepackage{epsfig,amsmath}
\usepackage{subfigure}
\usepackage{graphicx}
\usepackage{dcolumn}
\usepackage{stmaryrd}
\usepackage{mathrsfs}
\usepackage{pifont}
\usepackage{amsthm}
\usepackage{amssymb}
\usepackage{bm}
\usepackage{latexsym}
\usepackage[colorlinks=true,linkcolor=blue,citecolor=blue]{hyperref}
\usepackage{color}
\usepackage{epstopdf}

\begin{document}

\title{Generating NOON states in circuit QED using multi-photon resonance in the presence of counter-rotating interactions}

\author{Shi-fan Qi}
\affiliation{Zhejiang Province Key Laboratory of Quantum Technology and Device, Department of Physics, Zhejiang University, Hangzhou 310027, Zhejiang, China}

\author{Jun Jing}
\email{Email address: jingjun@zju.edu.cn}
\affiliation{Zhejiang Province Key Laboratory of Quantum Technology and Device, Department of Physics, Zhejiang University, Hangzhou 310027, Zhejiang, China}

\date{\today}

\begin{abstract}
The NOON states are valuable quantum resources, which have a wide range of applications in quantum communication, quantum metrology, and quantum information processing. Here we propose a fast, concise and reliable protocol for deterministically generating the NOON states of two resonators coupled to a single $\triangle$-type superconducting qutrit. In particular, we derive the effective Hamiltonians at the multi-photon resonances by virtue of the strong counter-rotating interaction between the resonator modes and the qutrit. Based on these crucial effective Hamiltonians, our protocol simplifies the previous ones using the single-photon resonance and consequently reduces the number of operations for state preparation. To test the robustness of this protocol, we analyze the effects from both the decoherence including dissipation and dephasing and the crosstalk of resonator modes on the state fidelity through a Lindblad master equation in the eigenstates of the full Hamiltonian.
\end{abstract}

\maketitle
\section{Introduction}\label{in}

Entanglement~\cite{qe} plays a key role in quantum technologies, such as quantum communication~\cite{qco}, quantum computing, and quantum information processing~\cite{in}. The entangled states serve as valuable resources especially for the quantum communication protocols, including quantum key distribution~\cite{kd}, quantum secret sharing~\cite{qss}, and quantum secure direct communication~\cite{qsc,qsc2}. Novel quantum platforms as well as fast protocols for preparing and measuring the entangled states have therefore been intensively pursued for a long time~\cite{ghz,mp,pm,bg}, and are still under active investigation.

Among various entangled states, the NOON states $(|N0\rangle+|0N\rangle)/\sqrt{2}$ with $N$ integer consist of two orthogonal component states in maximal superposition, which have been applied in quantum optical lithography~\cite{ql1,ql2}, quantum metrology~\cite{qm1,qm2} and quantum information processing~\cite{qco,qip}. To populate the Fock states with more photons in preparing the NOON state, it is of great helpful to have adjustable atomic level splitting to be selectively resonant with the desired process. Many protocols in the circuit-QED systems have thus been proposed upon the ability of level-manipulation in the artificial qubit, qutrit or qudit systems. The setup for NOON state-preparation in Ref.~\cite{noon1} consists of two superconducting resonators and a tunable qubit under classical driving. The operation number in that protocol is increased in a nonlinear way in terms of the number of the photons $N$. Although in Ref.~\cite{noon2}, only a linear number of operations with respect to $N$ are required for NOON-state preparation, the circuit on demand consists of more elements (three resonators and two qutrits) in the experiment~\cite{noon3}. The protocol in Ref.~\cite{noon4} adopts a setup consisting of one superconducting transmon qutrit and two resonators, which demands $2N$ operations to generate a NOON state with $N$ photons. It has been shortly improved by Ref.~\cite{noon5}, which demands only $N+1$ operations while using a four-level superconducting flux device and two resonators. Recently, an efficient protocol~\cite{noon6} for generating a ``double" NOON state $(|NN00\rangle+|00NN\rangle)/\sqrt{2}$ was proposed, which consists of $N+2$ operations. However, it employs five resonators and two superconducting flux qutrits (the latter are initialized as a Bell state). Therefore, a trade-off has to be made between the complexity of the devices (as well as their initial states) and the number of running steps towards the NOON states.

In this work, we propose a fast protocol for generating the NOON states of two resonators strongly coupled to a superconducting qutrit. The strong or ultrastrong coupling opens a door to study the physics of virtual processes governed by the interaction Hamiltonian and leads to many interesting phenomena and applications~\cite{ip,ip2,ip3,ip4,ip5,ip6,ip7,ip8,ip9,ip10}. With respect to the ratio of coupling strength and the characteristic frequency in the Rabi model, $g/\omega\simeq0.1$ is a rough starting point of the strong-coupling regime~\cite{ip9}. In the strongly coupled qubit-cavity systems, Rabi Hamiltonians~\cite{sc,sc2} have been applied in the protocol~\cite{bg} to prepare the Bell states and the GHZ states. Without the counter-rotating terms, it is impossible to achieve a desired effective Hamiltonian connecting states with no conservation of the number of photons or excitons from the full Hamiltonian. Here we extend the ideas in Ref.~\cite{sc,sc2} to prepare the NOON states in light of the feasible capabilities of manipulating the artificial atoms and atom-photon interactions in the circuit-QED systems.

Around the specific multi-photon resonance points, we extract the effective Hamiltonians for the interested system using the perturbation theory~\cite{ip8,bg,ql}. At such a point, a selective pair of levels of the $\triangle$-type qutrit are multi-photon resonant with the corresponding resonator, while the remaining level and resonator do not interfere with the preceding resonance as well as the relevant Hilbert subspace. Then our model of the $\triangle$-type qutrit coupled to two resonators can be regarded as a ``parallel'' Rabi model. The separability of the two subspaces and the valid regime of the effective Hamiltonians have been verified by the standard numerical diagonalization of the full Hamiltonian. The NOON states can be deterministically realized in the effective model. Regarding the multi-photon resonance based on the strong or ultrastrong atom-photon interaction and the controllability of the atomic frequencies, our proposal can be implemented in the circuit-QED systems~\cite{sq,sq2,sq3,sq4}.

The rest of the work is organized as follows. In Sec.~\ref{model}, we introduce the full Hamiltonian. We numerically discuss and analytically explain the multi-photon resonances arising from the counter-rotating interaction in the strong-coupling regime. The double-photon and triple-photon resonances are discussed in Sec.~\ref{twophoton} and Sec.~\ref{threephoton}, respectively. The detailed derivation of the effective Hamiltonians can be found in Appendices~\ref{appa} and \ref{appb}. In Sec.~\ref{fide}, we analyze the decoherence effect from the external environment on the state fidelity by a Lindblad master equation written in the eigenbases of the full Hamiltonian. In Sec.~\ref{pp}, we provide a detailed procedure for the NOON state preparation with an arbitrary $N$ and then consider the effect of the inter-resonator coupling on their fidelities. In Sec.~\ref{dc}, we conclude the work with a brief discussion.

\section{Description of the model}\label{model}

\subsection{Model Hamiltonian}\label{hm}

The model we investigated is a general Rabi model describing a superconducting $\triangle$-type qutrit coupled to two resonators (labelled $a$ and $b$) in both longitudinal and transversal directions. The three levels of the $\triangle$-type qutrit are labelled by $|g\rangle$, $|e\rangle$, and $|f\rangle$, representing the ground state, the mediate state and the highest-level state, respectively. The system Hamiltonian $(\hbar\equiv1)$ can be written as
\begin{equation}\label{noonmodel}
\begin{aligned}
H&=H_0+V, \\
H_0&=\omega_a a^{\dag}a+\omega_b b^{\dag}b+\omega_{eg}|e\rangle\langle e|+\omega_{fg}|f\rangle\langle f|,\\
V&=[g_{eg}^a(a^{\dag}+a)+g_{eg}^b(b+b^{\dag})](\sigma_x^{eg}\cos\theta+\sigma_z^{eg}\sin\theta)\\
&+[g_{fg}^a(a^{\dag}+a)+g_{fg}^b(b+b^{\dag})](\sigma_x^{fg}\cos\theta+\sigma_z^{fg}\sin\theta)\\
&+[g_{fe}^a(a^{\dag}+a)+g_{fe}^b(b+b^{\dag})](\sigma_x^{fe}\cos\theta+\sigma_z^{fe}\sin\theta).
\end{aligned}
\end{equation}
Here $\omega_a$ and $\omega_b$ are the transition frequencies of the resonators $a$ and $b$, respectively. $\omega_{jk}\equiv\omega_j-\omega_k$ with $j,k\in\{g, e, f\}$ represents the level spacing between $|j\rangle$ and $|k\rangle$. The ground state energy is set as zero. $g^{n}_{jk}$ with $n\in\{a, b\}$ and $j,k\in\{g, e, f\}$ is the coupling strength between the resonator $n$ and the specific pair of levels $j$ and $k$. The coupling operators of the qutrit are defined as $\sigma_x^{jk}\equiv|j\rangle\langle k|+|k\rangle\langle j|$ and $\sigma_z^{jk}\equiv|j\rangle\langle j|-|k\rangle\langle k|$, respectively. The angle $\theta$ parameterizes the amount of the longitudinal and the transversal coupling between the qutrit and the resonators, which is adjustable and independent from the choice of the coupling strength $g^{n}_{jk}$. The arbitrary mixture of the longitudinal and the transversal couplings has been realized in the circuit-QED experiments~\cite{ip3,cq}.

To simplify the discussion but with no loss of generality, we let $g_{eg}^{n}=g_{fg}^n=g_{fe}^n=g_n$ with $n=a,b$. Thus, the interaction Hamiltonian $V$ in Eq.~\eqref{noonmodel} can be written as
\begin{equation}\label{V}
\begin{aligned}
V&=[g_a(a^{\dag}+a)+g_b(b+b^{\dag})]\big(\sigma_x^{eg}\cos\theta+\sigma_z^{eg}\sin\theta\\
&+\sigma_x^{fg}\cos\theta+\sigma_z^{fg}\sin\theta+\sigma_x^{fe}\cos\theta+\sigma_z^{fe}\sin\theta\big).
\end{aligned}
\end{equation}
The rest part of this work is based on the Hamiltonian~(\ref{V}). We would apply the standard perturbation theory to obtain the effective Hamiltonians in charge of the desired Rabi oscillations, by which we can construct the NOON states. Also for each effective Hamiltonian, a pair of effective coupling strength and energy shift, which are consistent with each other to the leading-order of $g_a$ or $g_b$, can be determined. The energy shifts as well as the relevant level spacing for the multi-photon resonances are used in the comparison between the numerical and analytical results.

\subsection{Two-photon resonance}\label{twophoton}

\begin{figure}[htbp]
\centering
\includegraphics[width=0.45\textwidth]{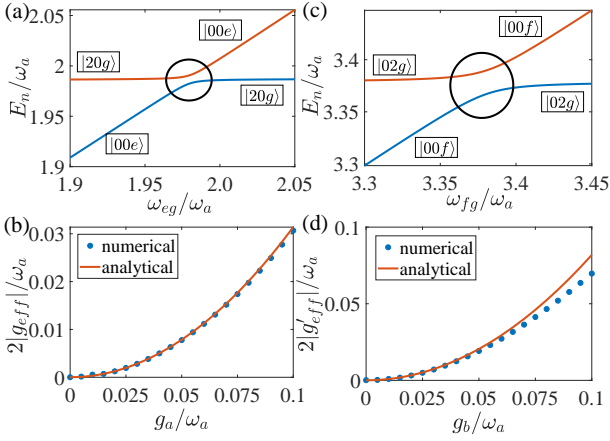}
\caption{(Color online) (a) and (c) Energy-levels and avoided level-crossings for the two-photon resonances, normalized by $\omega_a$ and plotted as a function of the transition frequency $\omega_{eg}$ and $\omega_{fg}$, respectively. The avoided level-crossings of the eigenstates are distinguished in the dark circles. (b) and (d) Comparison between the numerically calculated normalized effective coupling strengths $|g_{eff}|/\omega_a$ and $|g_{eff}'|/\omega_a$ (blue dots) and the corresponding analytical results in Eqs.~\eqref{ge} and \eqref{ges} from the second-order perturbation theory (orange solid line), respectively. Here we fix $\omega_b=1.7\omega_a$, $\theta=\pi/6$. For (a) and (b), $\omega_{fg}=2\omega_b$ and $g_b=0.05\omega_a$; for (c) and (d), $\omega_{eg}=2\omega_a$ and $g_a=0.05\omega_a$.}\label{two}
\end{figure}

To show the basic mechanism for our proposal about the NOON-state generation by the double-photon resonance, we plot in Fig.~\ref{two} the avoided level-crossings of the system as a function of the qutrit frequencies. The eigenvalues $\{E_n\}$ and the eigenstates of the full Hamiltonian~\eqref{noonmodel} are obtained by standard numerical diagonalization method in a truncated Hilbert space. A sufficiently large number of energy eigenstates have been used to ensure that the treatment here is not significantly affected by the truncation.

In Fig.~\ref{two}(a), we see that an avoided level-crossing (distinguished in the dark circle) occurs between two eigenstates of the full Hamiltonian when the level spacing $\omega_{eg}$ approaches $2\omega_a$. It demonstrates a two-photon resonance, in which two photons of mode-$a$ can be simultaneously created by the atomic transition from level $|e\rangle$ to level $|g\rangle$, i.e., $|00e\rangle\rightarrow|20g\rangle$. And inversely, two photons of mode-$a$ can be simultaneously annihilated by the atomic transition from $|g\rangle$ to $|e\rangle$, i.e., $|20g\rangle\rightarrow|00e\rangle$. In fact, when $\omega_{eg}\approx2\omega_a$, $|00e\rangle$ and $|20g\rangle$ are nearly degenerate and they become the main component of the system eigenstates with $E_n\approx2\omega_a$. At the avoided-crossing point (the middle point in the dark circle), $|E_n\rangle\approx(|00e\rangle\pm|20g\rangle)/\sqrt{2}$. The lower eigenvector approaches $|00e\rangle$ ($|20g\rangle$) at the red (blue) far-off-resonant end and the upper eigenvector does the other way around. The existence of $V$ in either Eq.~(\ref{noonmodel}) or Eq.~(\ref{V}) lifts the degeneracy of the two eigenstates and renders a strong Rabi-oscillation between them. This phenomenon can also be well-explained by the effective Hamiltonian due to the second-order process involving both the longitudinal and transversal couplings in Eq.~\eqref{noonmodel}, provided that $g_a, g_b\ll\omega_a, \omega_b, |\omega_b-\omega_a|$. The detailed derivation can be found in Appendix~\ref{appa}. The effective Hamiltonian is found to be
\begin{equation}\label{Heff}
H_{eff}=g_{eff}\left(|00e\rangle\langle 20g|+|20g\rangle\langle 00e|\right).
\end{equation}
Here the coupling strength
\begin{equation}\label{ge}
g_{eff}=-\sqrt{2}g^2_a\left[\frac{\sin(2\theta)}{\omega_a}+\frac{\cos^2\theta}{\omega_{fg}-\omega_a}\right]
\end{equation}
is in the same order as the distinction $\delta$ of the avoided level-crossing point from $2\omega_a$. At this point, $\omega_{eg}=2\omega_a+\delta$, where $\delta$ is
\begin{equation}\label{de}
\begin{aligned}
\delta&=-g^2_a\cos^2\theta\left(\frac{4}{\omega_a}+\frac{1}{\omega_{fg}-\omega_a}+\frac{3}{\omega_{fg}+\omega_a}\right)\\
&-g^2_b\cos^2\theta\bigg(\frac{1}{2\omega_a+\omega_b}+\frac{1}{\omega_{fg}+\omega_b}+\frac{1}{2\omega_a-\omega_b}\\
&+\frac{1}{2\omega_{a}-\omega_b-\omega_{fg}}\bigg)-\frac{4g^2_a\sin^2\theta}{\omega_a}-\frac{4g^2_b\sin^2\theta}{\omega_b}.
\end{aligned}
\end{equation}
The energy-splitting of the two eigenstates at the avoided level-crossing point is $2|g_{eff}|$, which can be evaluated by the numerical simulation over the whole Hilbert space. The comparison of $|g_{eff}|$ between the analytical~\eqref{ge} and numerical~\eqref{noonmodel} results are shown in Fig.~\ref{two}(b) as a function of normalized coupling strength $g_a/\omega_a$. The orange solid line is the analytical result from the effective Hamiltonian~\eqref{ge} and the blue dots are the results from the numerical diagonalization of the full Hamiltonian~\eqref{noonmodel}. One can see that the effective Hamiltonian yields perfect results for normalized interaction strengths $g_a/\omega_a\leq0.095$. For an even larger $g_a$, higher-orders contribution needs to capture all the effects from the interaction Hamiltonian modifying the eigenstates of the bare system.

Figures~\ref{two}(c) and \ref{two}(d) demonstrate the avoided level-crossing at the double-photon resonance occurring in the subspace spanned by $\{|00f\rangle, |02g\rangle\}$ (two hybridized states by the atomic levels $|f\rangle$, $|g\rangle$ and mode-$b$) when the level spacing $\omega_{fg}$ approaches $2\omega_b$. Here we choose $\omega_b=1.7\omega_a$ to avoid the unnecessary mixture of the demanding near-degenerate eigenstates or subspaces. Followed by a similar perturbative derivation (see Appendix~\ref{appa}), the effective Hamiltonian is found to be
\begin{equation}\label{Heffs}
H_{eff}'=g_{eff}'(|00f\rangle\langle 02g|+|02g\rangle\langle 00f|),
\end{equation}
where
\begin{equation}\label{ges}
g_{eff}'=-\sqrt{2}g^2_b\left[\frac{2\sin(2\theta)}{\omega_b}+\frac{\cos^2\theta}{\omega_{eg}-\omega_b}\right].
\end{equation}
To the second order of $g_b$, the energy shift is evaluated by
\begin{equation}\label{De}
\begin{aligned}
\Delta&=\omega_{fg}-2\omega_b \\
&=-g^2_b\cos^2\theta\left(\frac{4}{\omega_b}+\frac{1}{\omega_{eg}-\omega_b}+\frac{3}{\omega_{eg}+\omega_b}\right)\\
&-g^2_a\cos^2\theta\bigg(\frac{1}{\omega_{eg}+\omega_a}+\frac{1}{2\omega_b+\omega_a}+\frac{1}{2\omega_b-\omega_a}\\
&+\frac{1}{2\omega_b-\omega_a-\omega_{eg}}\bigg).
\end{aligned}
\end{equation}
Here the normalized effective coupling strength for the double-photon resonance is demonstrated in Fig.~\ref{two}(d). Similar to Fig.~\ref{two}(b), one can also roughly estimate the valid range of the second-order perturbative Hamiltonian through the comparison between the analytical result by Eq.~\eqref{ges} and the numerical evaluation by the Hamiltonian~\eqref{noonmodel}. The two results match with each other for normalized atom-photon interaction strength $g_b/\omega_a\leq0.065$, the upper bound of which is still of a strong-coupling regime, and depends on the choice of the other parameters, such as $g_a$, $\omega_b$, and $\omega_{eg}$.

\begin{figure}[htbp]
\centering
\includegraphics[width=0.4\textwidth]{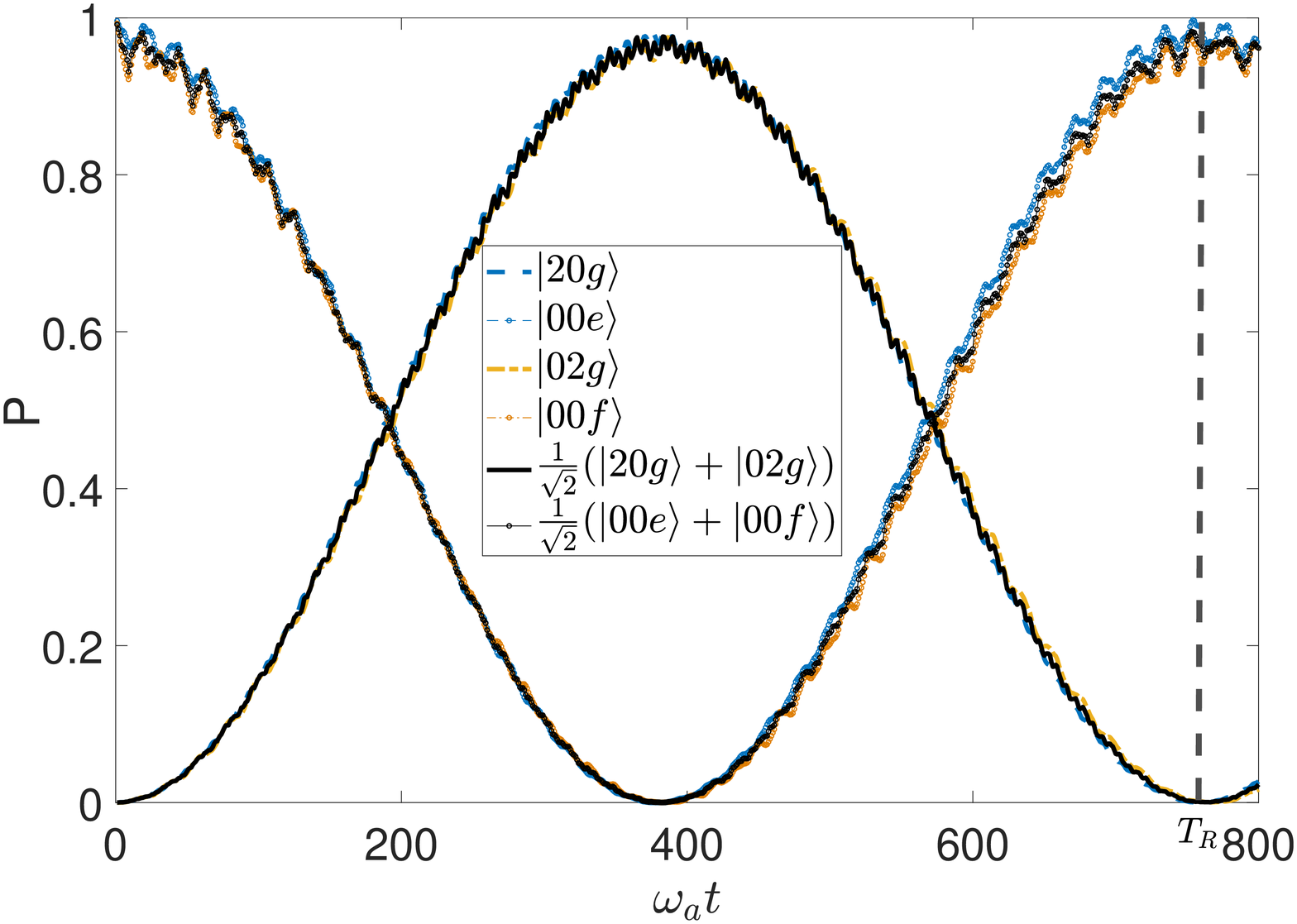}
\caption{(Color online) Three groups of Rabi oscillations in one period. (1) The blue dashed line and the blue dashed line with dots represent the populations of state $|20g\rangle$ and $|00e\rangle$, respectively (the initial state is $|00e\rangle$). (2) The orange dash-dotted line and the orange dashed line with dots represent the state $|02g\rangle$ and $|00f\rangle$, respectively (the initial state is $|00f\rangle$). (3) The dark solid line and the dark solid line with dots represent the state $(|20g\rangle+|02g\rangle)/\sqrt{2}$ and $(|00e\rangle+|00f\rangle)/\sqrt{2}$, respectively (the initial state is $(|00e\rangle+|00f\rangle)/\sqrt{2}$). The parameters are chosen as $\omega_b=1.7\omega_a$, $g_a=0.05\omega_a$ and $\theta=\frac{\pi}{6}$. The vertical grey dashed line represents the analytical result. }\label{twonoon}
\end{figure}

The two effective Hamiltonians~(\ref{Heff}) and (\ref{Heffs}) are independent with each other and can be simultaneously constructed provided $\omega_{eg}=2\omega_a+\delta$ and $\omega_{fg}=2\omega_b+\Delta$. When the whole system is initialled at the $|00e\rangle$, i.e., the two resonators are in the vacuum state and the qutrit is at $|e\rangle$, a completed Rabi oscillation between $|00e\rangle$ and $|20g\rangle$ can then be accurately realized with a period $T_R=\pi/|g_{eff}|$. Similarly, when the whole system is initialled as $|00f\rangle$, a completed Rabi oscillation between $|00f\rangle$ and $|02g\rangle$ can also be observed with a period $T'_R=\pi/|g'_{eff}|$. Thus if the system is initialled as the symmetrical superposed state $(|00e\rangle+|00f\rangle)/\sqrt{2}$ and $T_R=T'_R$, it will evolve into $(|20g\rangle+|02g\rangle)/\sqrt{2}=(|20\rangle+|02\rangle)/\sqrt{2}\otimes|g\rangle$ at $t=T_R/2$. For the resonator modes, they are now prepared as the two-photon NOON state. In Fig.~\ref{twonoon}, we plot all these Rabi oscillations in one period under the constrain $T_R=T'_R$. With the parameters we chosen, it is found that $\omega_aT_R=\pi/|g_{eff}|\approx754$ by Eq.~(\ref{ge}), which is confirmed by the numerical simulation.

It is noteworthy to point that if $T_R\neq T'_R$, we can firstly let the initial state to experience a half Rabi-oscillation driven by Eq.~(\ref{Heff}); then use a microwave $\pi/2$ pulse to realize $|e\rangle\to|g\rangle$; and finally control the system to undergo a half Rabi-oscillation driven by Eq.~(\ref{Heffs}). In the end, the NOON state $(|20\rangle+|02\rangle)/\sqrt{2}$ emerges, consuming more time and resource than that in case of $T_R=T'_R$. The order of these two half-Rabi-oscillations can be exchanged if the $\pi/2$ pulse is performed between $|f\rangle$ and $|g\rangle$.

With the effective Hamiltonians~(\ref{Heff}) and (\ref{Heffs}), one can generate a double-photon NOON state from the vacuum state of modes $a$ and $b$. This protocol can be straightforwardly extended to generate a $(n+2)$-photon NOON state when the system is in the state $(|n0e\rangle+|0nf\rangle)/\sqrt{2}$. In the subspace spanned by $\{|n0e\rangle, |(n+2)0g\rangle\}$, one can construct an effective Hamiltonian $H_{eff}=g_{eff}(n)[|n0e\rangle\langle(n+2)0g|+|(n+2)0g\rangle\langle n0e|]$ through a derivation similar to Appendix~\ref{appa}. Here the effective coupling strength is $n$-dependent:
\begin{equation}\label{geffn12}
\begin{aligned}
g_{eff}(n)=\sqrt{\frac{(n+1)(n+2)}{2}}g_{eff},
\end{aligned}
\end{equation}
where $g_{eff}$ is found in Eq.~(\ref{ge}) and the energy shift is also $n$-dependent:
\begin{equation}\label{den}
\delta_n=\delta-ng^2_a\cos^2\theta\left(\frac{8}{3\omega_a}+\frac{1}{\omega_{fg}+\omega_a}-\frac{1}
{\omega_{fg}-3\omega_a}\right).
\end{equation}
While in the subspace spanned by $\{|0nf\rangle, |0(n+2)g\rangle\}$, one can find
\begin{equation}\label{geffn34}
\begin{aligned}
g_{eff}'(n)=\sqrt{\frac{(n+1)(n+2)}{2}}g'_{eff},
\end{aligned}
\end{equation}
and
\begin{equation}\label{deln}
\Delta_n=\Delta-ng^2_b\cos^2\theta\left(\frac{8}{3\omega_b}+\frac{1}{\omega_{eg}+\omega_b}-\frac{1}
{\omega_{eg}-3\omega_b}\right).
\end{equation}
Note $\delta$ in Eq.~(\ref{den}) and $\Delta$ in Eq.~(\ref{deln}) represent the energy shifts in Eqs.~(\ref{de}) and (\ref{De}), respectively. Roughly, the magnitude of the effective coupling strength scales linearly with the initial photon number of the resonators, which accordingly reduces the time for state-preparation.

\subsection{Three-photon resonance}\label{threephoton}

In this subsection, we provide a protocol to achieve the three-photon resonances, which is based on Eq.~\eqref{V} with $\theta=0$. The Hamiltonian with a nonvanishing $\theta$ is of course available for the same target. Nevertheless the relevant protocol raises more treatments in mathematics yet no extra physics. Now the full Hamiltonian can be written as
\begin{equation}\label{threemodel}
\begin{aligned}
H&=H_0+V,\\
H_0&=\omega_{eg}|e\rangle\langle e|+\omega_{fg}|f\rangle\langle f|+\omega_a a^{\dag}a+\omega_b b^{\dag}b,\\
V&=[g_a(a^{\dag}+a)+g_b(b+b^{\dag})](\sigma_x^{eg}+\sigma_x^{fg}+\sigma_x^{fe}).
\end{aligned}
\end{equation}

\begin{figure}[htbp]
\centering
\includegraphics[width=0.45\textwidth]{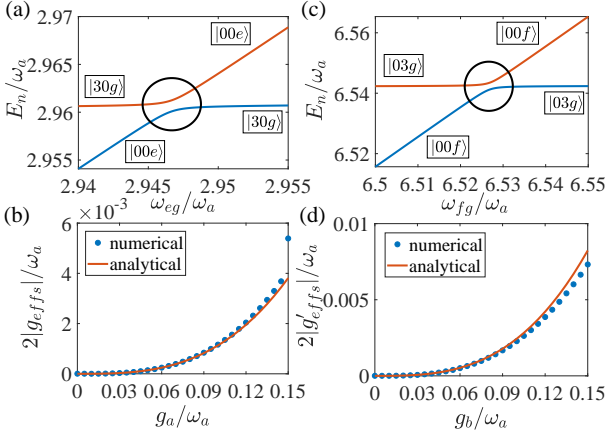}
\caption{(Color online) (a) and (c) Energy-levels and avoided level-crossings for the three-photon resonances, normalized by $\omega_a$ and plotted as a function of the transition frequency $\omega_{eg}$ and $\omega_{fg}$, respectively. (b) and (d) Comparison between the numerically calculated normalized effective coupling strengths $|g_{effs}|/\omega_a$ and $|g_{effs}'|/\omega_a$ (blue points) and the corresponding analytical results~\eqref{eg} and~\eqref{egs} (orange solid line), respectively. Here we fix $\omega_b=2.2\omega_a$. For (a) and (b), $\omega_{fg}=3\omega_b$ and $g_b=0.1\omega_a$; for (c) and (d), $\omega_{eg}=3\omega_a$ and $g_a=0.1\omega_a$.}\label{three}
\end{figure}

Analog to Sec.~\ref{twophoton}, here in Fig.~\ref{three}(a) and Fig.~\ref{three}(c), we describe two avoided level-crossings required for three-photon resonances in the respective subspaces. In Fig.~\ref{three}(b) and Fig.~\ref{three}(d), we justified the corresponding effective Hamiltonians by comparing the effective coupling strengthes based on analytical derivation to those obtained from the numerical results at the avoided level-crossing points. Particularly, Fig.~\ref{three}(a) demonstrates the avoided level-crossing (distinguished in dark circle) when $\omega_{eg}$ approaches $3\omega_a$ in the subspace spanned by $\{|00e\rangle, |30g\rangle\}$. Through a straightforward derivation in Appendix~\ref{appb}, the effective Hamiltonian is found to be
\begin{equation}\label{effH}
H_{effs}=g_{effs}\left(|00e\rangle\langle 30g|+|30g\rangle\langle 00e|\right),
\end{equation}
where
\begin{equation}\label{eg}
g_{effs}=-\frac{\sqrt{6}g^3_a}{2\omega_a}\left(\frac{1}{2\omega_a}
-\frac{1}{\omega_{fg}-2\omega_a}+\frac{1}{\omega_{fg}-\omega_a}\right).
\end{equation}
The leading-order correction of $\omega_{eg}$ to $3\omega_a$ can be obtained as
\begin{equation}\label{delta}
\begin{aligned}
\delta_s&=\omega_{eg}-3\omega_a \\
&=-g^2_a\left(\frac{3}{\omega_a}+\frac{3}{\omega_{fg}-\omega_a}+\frac{4}{\omega_{fg}+\omega_a}+\frac{1}{2\omega_a-\omega_{fg}}\right)\\
&-g^2_b\bigg(\frac{1}{3\omega_a+\omega_b}+\frac{1}{\omega_b+\omega_{fg}}+\frac{1}{3\omega_a-\omega_b}\\
&+\frac{1}{3\omega_a-\omega_b-\omega_{fg}}\bigg).
\end{aligned}
\end{equation}
In Fig~\ref{three}(b), one can observe that the analytical result of the effective energy splitting~(\ref{eg}) remains perfect for the normalized interaction strengths $g_a/\omega_a\leq0.14$ in comparison with the numerical result obtained from Eq.~\eqref{threemodel}.

Figure~\ref{three}(c) demonstrates the avoided level-crossing occurring in the subspace spanned by $\{|00f\rangle, |03g\rangle\}$ when the level spacing $\omega_{fg}$ approaches $3\omega_b$. Followed by a similar perturbative derivation as the preceding case (see Appendix~\ref{appb}), the effective Hamiltonian is found to be
\begin{equation}\label{effHs}
H'_{effs}=g_{effs}'\left(|00f\rangle\langle 03g|+|03g\rangle\langle 00f|\right),
\end{equation}
where the effective coupling strength
\begin{equation}\label{egs}
g_{effs}'=-\frac{\sqrt{6}g^3_b}{2\omega_b}\left(\frac{1}{2\omega_b}
-\frac{1}{\omega_{eg}-2\omega_b}+\frac{1}{\omega_{eg}-\omega_b}\right).
\end{equation}
The leading-order correction of the avoided level-crossing point, i.e., $\Delta_s=\omega_{fg}-3\omega_b$, can be obtained as
\begin{equation}\label{Delta}
\begin{aligned}
\Delta_s&=-g^2_a\bigg(\frac{1}{3\omega_b+\omega_a}+\frac{1}{\omega_a+\omega_{eg}}+\frac{1}{3\omega_b-\omega_a-\omega_{eg}}\\
&+\frac{1}{3\omega_b-\omega_a}\bigg)-g^2_b\bigg(\frac{1}{2\omega_b-\omega_{eg}}+\frac{3}{\omega_b}+\frac{3}{\omega_{eg}-\omega_b}\\
&+\frac{4}{\omega_{eg}+\omega_b}\bigg).
\end{aligned}
\end{equation}
And then we present both the analytical result in Eq.~\eqref{egs} and the numerical result with full Hamiltonian in Fig.~\ref{three}(d) to verify the regime in which they can match with each other. It is found that when $g_b/\omega_a\leq0.13$, the third-order perturbation remains as a good approximation. Although the upper bound of the coupling strength depends on the choice of parameters, it enters the strong-coupling regime.

\begin{figure}[htbp]
\centering
\includegraphics[width=0.4\textwidth]{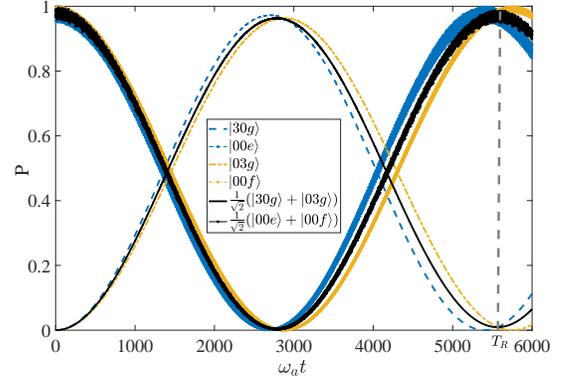}
\caption{(Color online) Three groups of Rabi oscillations. (1) The blue dashed line and the blue dashed line with dots represent the state $|30g\rangle$ and $|00e\rangle$, respectively (the initial state is $|00e\rangle$). (2) The orange dash-dotted line and the orange dashed line with dots represent the state $|03g\rangle$ and $|00f\rangle$, respectively (the initial state is $|00f\rangle$). (3) The dark solid line and the dark solid line with dots represent the state $(|30g\rangle+|03g\rangle)/\sqrt{2}$ and $(|00e\rangle+|00f\rangle)/\sqrt{2}$, respectively (the initial state is $(|00e\rangle+|00f\rangle)/\sqrt{2}$). Here $\omega_b=2.2\omega_a$ and $g_a=0.1\omega_a$.}\label{threenoon}
\end{figure}

As long as the energy levels of the artificial atom are adjusted to meet both $\omega_{eg}=3\omega_a+\delta_s$ and $\omega_{fg}=3\omega_b+\Delta_s$, the two effective Hamiltonians~(\ref{effH}) and (\ref{effHs}) can then be simultaneously constructed. Consequently, when the whole system is initialled as an arbitrary superposed state over $|00e\rangle$ and $|00f\rangle$ and $|g_{effs}|=|g'_{effs}|$, a completed Rabi oscillation occuring between the initial state and the corresponding superposed state over $|30g\rangle$ and $|03g\rangle$ with period $T_R=\pi/|g_{effs}|$. Thus if the system is initialled as $(|00e\rangle+|00f\rangle)/\sqrt{2}$, it will completely transform to $(|30g\rangle+|03g\rangle)/\sqrt{2}=(|30\rangle+|03\rangle)/\sqrt{2}\otimes|g\rangle$ at $t=T_R/2$. At this moment, the cavity modes are prepared as the three-photon NOON state. In Fig.~\ref{threenoon}, we plot the relevant Rabi oscillations in one period under certain parameters by exact numerical evaluations. With the chosen parameters, it is found that $\omega_aT_R=\pi/|g_{effs}|\approx 5562$ by Eq.~(\ref{eg}), which is confirmed by numerical simulation (see the vertical dashed grey line).

\section{Fidelity Analysis}\label{fide}

As shown in Figs.~\ref{twonoon} and \ref{threenoon}, the initial state $|00\rangle\otimes(|e\rangle+|f\rangle)/\sqrt{2}$ can be transformed to $(|20\rangle+|02\rangle)/\sqrt{2}\otimes|g\rangle$ and $(|30\rangle+|03\rangle)/\sqrt{2}\otimes|g\rangle$ at the respective multi-photon resonances, by undergoing a half Rabi oscillation. The states of the two modes are then the target NOON states with $N=2$ or $N=3$. Yet the whole system cannot be isolated from the surrounding environment. The target NOON state will be damaged by the influence from cavity mode damping and atomic decay and dephasing. In this section, the fidelity of the prepared state is studied based on the master equation approach. By applying the standard Markovian approximation and tracing out the degrees of freedom of external environment (assumed to be at the vacuum state), we arrive at the master equation~\cite{sc,me1,me2} for the density-matrix operator $\rho(t)$ of the whole system consisting of the two resonators and the artificial three-level atom,
\begin{equation}\label{master}
\begin{aligned}
\dot{\rho}(t)&=-i[H_{diag}, \rho(t)]+\kappa_a\mathcal{L}[X_a]\rho(t)+\kappa_b\mathcal{L}[X_b]\rho(t)\\
&+\gamma_{eg}\mathcal{L}[S_{eg}]\rho(t)+\gamma_{fg}\mathcal{L}[S_{fg}]\rho(t)+\gamma_{fe}\mathcal{L}[S_{fe}]\rho(t)\\
&+\gamma_{e}\mathcal{L}[S_{e}]\rho(t)+\gamma_{f}\mathcal{L}[S_{f}]\rho(t).
\end{aligned}
\end{equation}
Here $H_{diag}$ indicates that the full Hamiltonian $H$ is now expressed by its eigenvectors $|E_n\rangle$'s. $\kappa_a$ ($\kappa_b$) is the decay constant of the resonator mode $a$ ($b$). $\gamma_{eg}$, $\gamma_{fg}$, and $\gamma_{fe}$ are respectively the energy relaxation constants associated with the transitions $|e\rangle\to|g\rangle$, $|f\rangle\to|g\rangle$ and $|f\rangle\to|e\rangle$. $\gamma_{e}$ ($\gamma_{f}$) is the dephasing constant of the level $|e\rangle$ ($|f\rangle$). The superoperator $\mathcal{L}$ is defined as
\begin{equation}\label{D}
\mathcal{L}[O]\rho=\frac{1}{2}(2O\rho O^{\dag}-O^{\dag}O\rho-\rho O^{\dag}).
\end{equation}
Here $O=X_a, X_b, S_{eg}, S_{fg}, S_{fe}, S_{e}, S_{f}$ are the dressed lowering operators, defined respectively in terms of their bare counterparts $o=a, b, \sigma^-_{eg}, \sigma^-_{fg}, \sigma^-_{fe}, |e\rangle\langle e|, |f\rangle\langle f|$ as
\begin{equation}\label{O}
O\equiv\sum_{E_n>E_m}\langle E_m|(o+o^{\dag})|E_n\rangle|E_m\rangle\langle E_n|.
\end{equation}
To simplify the discussion but with no loss of generality, we let $\kappa_a=\kappa_b=\gamma_{fg}=\gamma_{fe}=\gamma_{eg}=\gamma_{e}=\gamma_{f}=\gamma$.

\begin{figure}[htbp]
\centering
\includegraphics[width=0.4\textwidth]{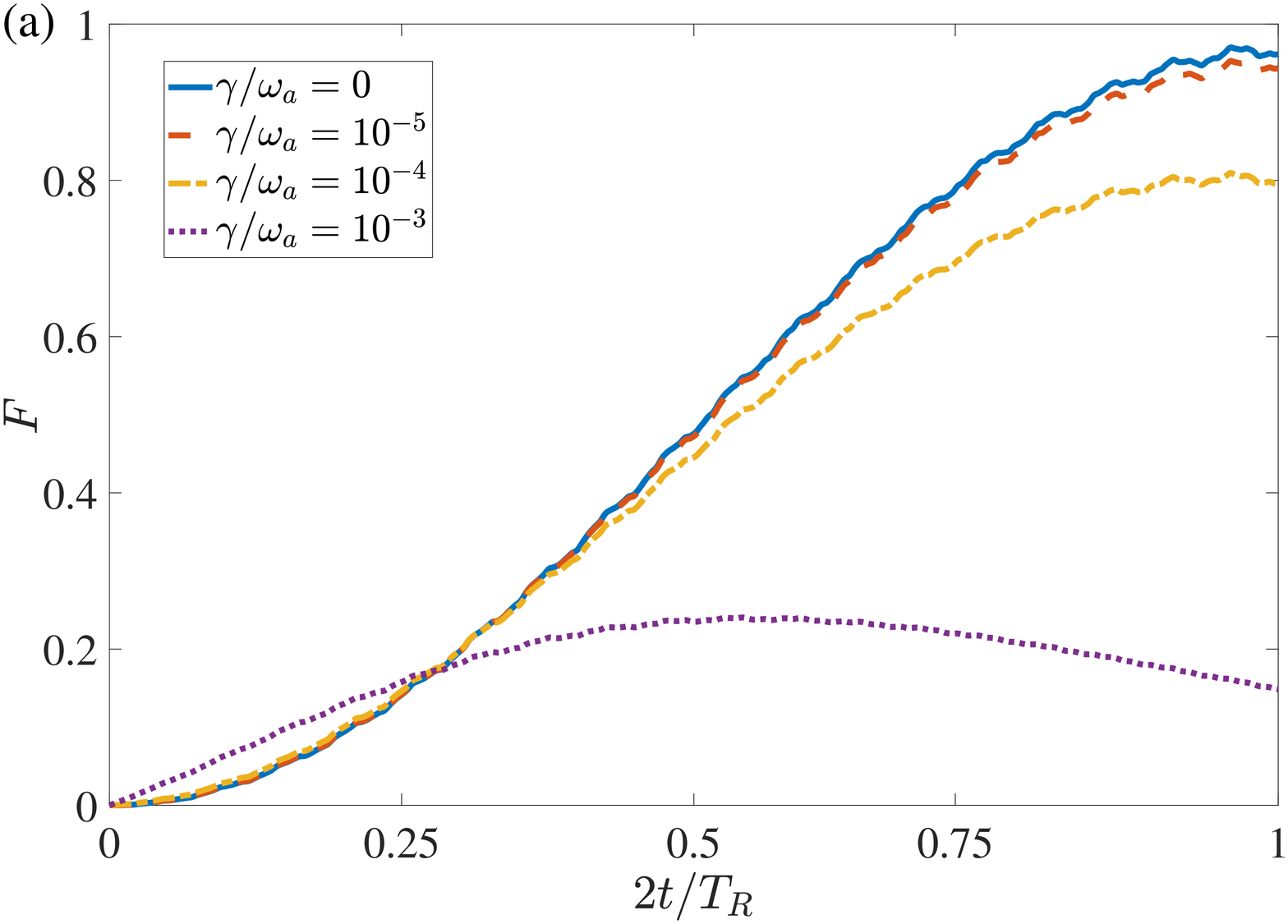}
\includegraphics[width=0.4\textwidth]{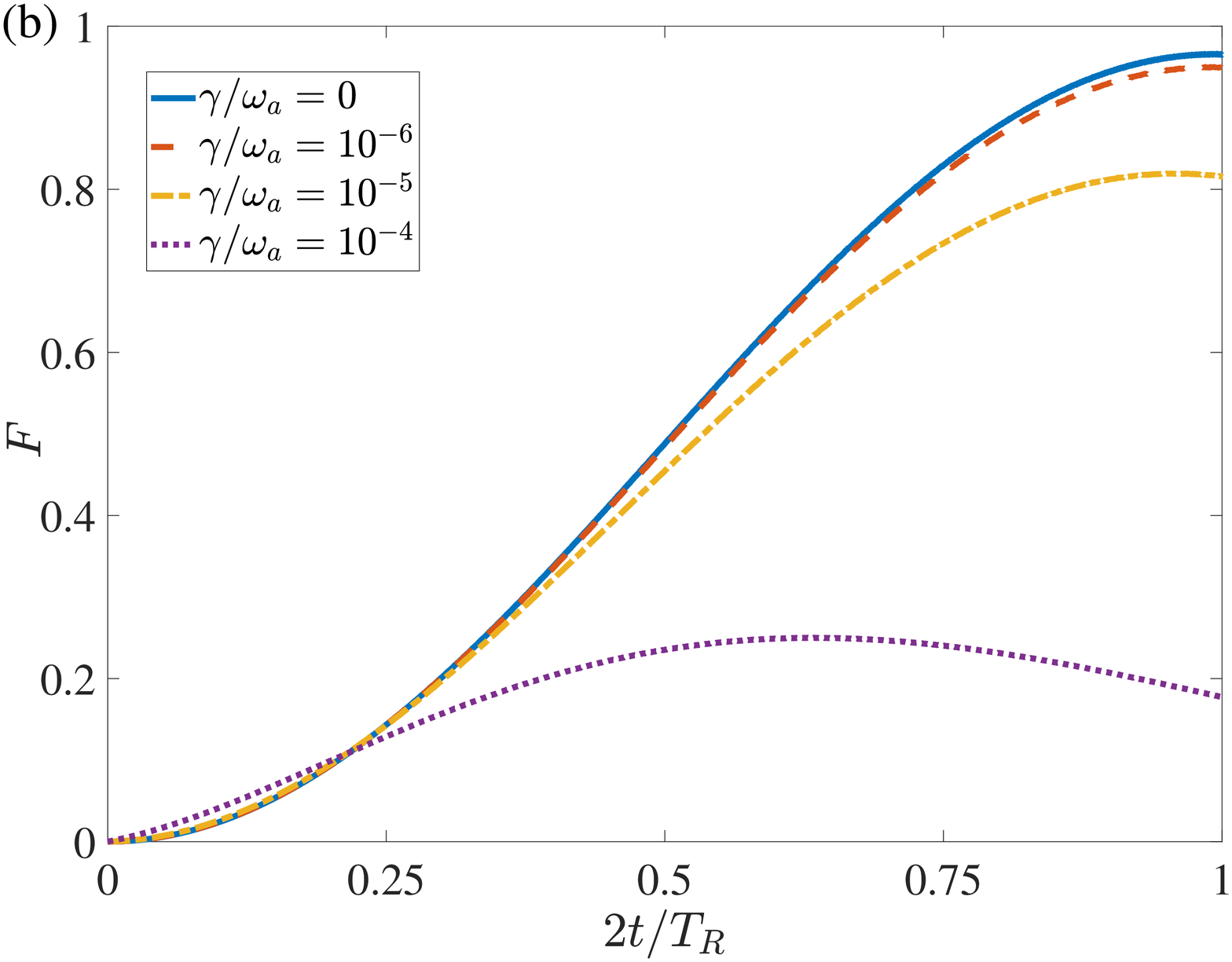}
\caption{(Color online) (a) and (b) Fidelities of the NOON state $(|20\rangle+|02\rangle)/\sqrt{2}$ and $(|30\rangle+|03\rangle)/\sqrt{2}$ prepared by our protocol given various decoherence rates, respectively. The parameters of (a) are set the same as those in Fig.~\ref{twonoon} and $T_R=\pi/|g_{eff}|$ is evaluated by Eq.~\eqref{ge}; the parameters of (b) are set the same as those in Fig.~\ref{threenoon} and $T_R=\pi/|g_{effs}|$ is evaluated by Eq.~\eqref{eg}.}\label{master23}
\end{figure}

The robustness of our protocols for generating $|\phi\rangle=(|N0\rangle+|0N\rangle)/\sqrt{2}$ upon the multi-photon resonance can be measured by the state-fidelity $F=\sqrt{\langle\phi|\rho(t)|\phi\rangle}$. Here $\rho(t)$ is numerically obtained by the master equation~(\ref{master}), whose parameters (the accurate value of level-spacings and evolution time) are determined in the derivation of the effective Hamiltonian.

In Fig.~\ref{master23}, we plot the fidelity under different decoherence rates $\gamma$. One can observe from Fig.~\ref{master23}(a) that the protocol of the double-photon resonance works rather well for $\gamma/\omega_a\leq10^{-5}$ (Note in recent experiments~\cite{fq,fq2,fq3,fq4}, the relative magnitude of the decoherence rates is about $10^{-6}\sim10^{-5}$), producing the desired NOON state with a fidelity of $96\%$, close to $97\%$ in the case with no decoherence. The fidelity $F$ maintains above $80\%$ even when $\gamma$ is enhanced to $10^{-4}\omega_a$. While for a larger decoherence rate $\gamma/\omega_a\leq10^{-3}$, the fidelity will drop below $50\%$. It is consistent with the fact that now $\gamma$ becomes comparable with the effective coupling strength of Eq.~(\ref{ge}). Note $|g_{eff}|/\omega_a\approx4\times10^{-3}$.

The fidelities for the protocol of the triple-photon resonance are shown in Fig.~\ref{master23}(b). It is clear that the protocol works well as long as $\gamma/\omega_a\leq10^{-6}$, producing the desired NOON state $|\phi\rangle=(|30\rangle+|03\rangle)/\sqrt{2}$ with a fidelity over $95\%$. The fidelity $F$ decreases to $82\%$ when $\gamma/\omega_a=10^{-5}$. And it is below $50\%$ when $\gamma/\omega_a=10^{-4}$, which is the same order as the effective coupling strength $|g_{effs}|/\omega_a\approx6\times10^{-4}$. It is also consistent with the fact that the effective coupling strength for the triple-photon process is one order weaker than that for the double-photon process.

\section{Preparing NOON states with multiple photons}\label{pp}

\begin{figure}[htbp]
\centering
\includegraphics[width=0.4\textwidth]{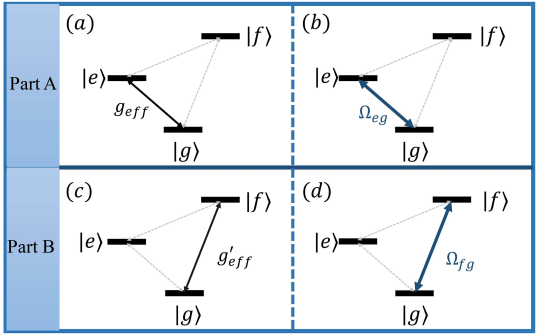}
\caption{(Color online) (a) and (c): Resonator $a$ ($b$) is tuned to be multi-photon resonant with the $|e\rangle \leftrightarrow |g\rangle$ ($|f\rangle \leftrightarrow |g\rangle$) transition and far-off multi-photon resonant with the $|f\rangle \leftrightarrow |g\rangle$ ($|e\rangle \leftrightarrow |g\rangle$) transition. Then the photon number of the mode is added while the qutrit is transformed to the ground state. (b) and (d): A driving pulse that is resonant with the $|e\rangle \leftrightarrow |g\rangle$ ($|f\rangle \leftrightarrow |g\rangle$) transition excites the qutrit back to the state $|e\rangle$ ($|f\rangle$).}\label{step}
\end{figure}

Suppose that the qutrit is initially in the state $|\psi\rangle=\frac{1}{\sqrt{2}}(|e\rangle+|f\rangle)$ and the two resonators are initially in the vacuum state $|00\rangle$. As shown in Fig.~\ref{step}, the procedure for generating the NOON states of the two resonators can be mainly divided into two parts: in part $A$ ($B$), the subspace $\{|e\rangle, |g\rangle\}$ ($\{|f\rangle, |g\rangle\}$) and the resonator mode $a$ ($b$) are operated by $(N-1)$ steps to populate the $|2(N-1)\rangle$ Fock state of the mode $a$ ($b$). Throughout the part $A$ ($B$), the irrelevant mode $b$ ($a$) is decoupled from the rest part of the whole system by tuning its frequency to the far-off-resonant regime. In the end, an operation by virtue of the effective dynamics involving both modes is performed to complete the whole protocol.

For the target NOON state with $2N$ photons ($N\geq1$), part-$A$ is divided into $N-1$ steps as following.

Step-$A_1$: Let mode-$a$ and the qutrit resonate with the transition $|00e\rangle \leftrightarrow |20g\rangle$, i.e., $\omega_{eg}=2\omega_a+\delta$, where $\delta$ is given by Eq.~\eqref{de}. Then as shown in Fig.~\ref{step}(a), the component $|00e\rangle$ in the initial state will become $|20g\rangle$ through the Rabi oscillation under the effective Hamiltonian~\eqref{Heff} after an evolution time $t_1=\pi/|2g_{eff}|$. Hence, the initial state $\Psi(0)=\frac{1}{\sqrt{2}}(|00e\rangle+|00f\rangle)$ of the whole system becomes
\begin{equation}\label{phit1}
\Psi(t_1)=\frac{1}{\sqrt{2}}(-i|20g\rangle+|00f\rangle).
\end{equation}
Then as shown in Fig.~\ref{step}(b), a microwave pulse of $\{\omega_{eg}, -\pi/2, \pi/(2\Omega_{eg})\}$ is applied to transform the state~\eqref{phit1} to
\begin{equation}\label{phit2}
\Psi(t_1+\tau)=\frac{1}{\sqrt{2}}(-i|20e\rangle+|00f\rangle),
\end{equation}
where $\omega_{eg}$ is the pulse frequency, $-\pi/2$ is a time-independent phase and $\tau\equiv\pi/(2\Omega_{eg})$ is the duration time of the pulse~\cite{noon4}. Here and below, we assume that Rabi frequency $\Omega_{eg}\gg|g_{eff}(n)|$ (Note $g_{eff}(0)=g_{eff}$), so that ideally the system evolution due to the qutrit-mode-$a$ interaction is negligible during $\tau$.

Step-$A_j$, $(j=2,3,\cdots,N-1)$: The level spacing $\omega_{eg}$ is tuned to be resonant with the transition $|(2j-2)0e\rangle \leftrightarrow |(2j)0g\rangle$, i.e., $\omega_{eg}=2\omega_a+\delta_{2j-2}$ according to Eq.~\eqref{den}. Through a similar Rabi oscillation as in step-$A_1$, $|(2j-2)0e\rangle$ is transformed to $-i|(2j)0g\rangle$ after $t_j=\pi/|2g_{eff}(2j-2)|$, where the effective coupling strength is given by Eq.~\eqref{geffn12}. $|(2j)0g\rangle$ is then further transformed to $|(2j)0e\rangle$ by a microwave pulse of $\{\omega_{eg}, -\pi/2, \pi/(2\Omega_{eg})\}$ pumping the state $|g\rangle$ back to $|e\rangle$. Thus after these operations, the state~\eqref{phit2} becomes
\begin{equation}\label{phitA}
\Psi(T_A)=\frac{1}{\sqrt{2}}\left[(-i)^{N-1}|(2N-2)0e\rangle+|00f\rangle\right],
\end{equation}
where $T_A=\sum^{N-1}_{j=1}t_j+(N-1)\tau$. At this moment, part-$A$ is completed. Then we focus on the $|00f\rangle \to |02g\rangle$ transition involving mode-$b$ as shown in Fig.~\ref{step}(c).

Step-$B_1$: Similar to step-$A_1$, the effective Hamiltonian~\eqref{Heffs} is used to realize a half Rabi transition from $|00f\rangle$ to $|02g\rangle$. After the interaction time $t'_1=\pi/|2g'_{eff}|$, where $g'_{eff}$ is given by Eq.~\eqref{ges}, the state~\eqref{phitA} becomes
\begin{equation}\label{phit1s}
\Psi(T_A+t'_1)=\frac{1}{\sqrt{2}}\left[(-i)^{N-1}|(2N-2)0e\rangle-i|02g\rangle\right].
\end{equation}
Then as shown in Fig.~\ref{step}(d), a microwave pulse of $\{\omega_{fg}, -\pi/2, \tau'\equiv\pi/(2\Omega_{fg})\}$ pumping $|g\rangle$ to $|f\rangle$, transforms the state~\eqref{phit1s} to
\begin{equation}\label{phit2s}
\Psi(T_A+t'_1+\tau')=\frac{1}{\sqrt{2}}\left[(-i)^{N-1}|(2N-2)0e\rangle-i|02f\rangle\right].
\end{equation}
Here and below, we assume $\Omega_{fg}\gg|g_{eff}'(n)|$ so that the interaction between the qutrit and mode-$b$ is negligible in the pulse duration time.

Step-$B_k$, $(k=2,3,...,N-1)$: Tune the level difference $\omega_{fg}$ to be resonant with the transition $|0(2k-2)f\rangle \leftrightarrow |0(2k)g\rangle$, namely, $\omega_{fg}=2\omega_b+\Delta_{2k-2}$ according to Eq.~\eqref{deln}. The system undergoes a similar Rabi oscillation as in step-$B_1$ with a evolution time set as $t_k'=\pi/|2g_{eff}'(2k-2)|$ due to Eq.~\eqref{geffn34}, after which the basis $|0(2k-2)f\rangle$ moves to $-i|0(2k)g\rangle$. It then moves to $-i|0(2k)f\rangle$ by a microwave pulse of $\{\omega_{fg},-\pi/2,\pi/(2\Omega_{fg})\}$ pumping $|g\rangle$ to $|f\rangle$. Thus after these steps, the state~\eqref{phit2s} becomes
\begin{equation}\label{phitB}
\begin{aligned}
\psi(T_A+T_B)&=\frac{1}{\sqrt{2}}\big[(-i)^{N-1}|(2N-2)0e\rangle\\
&+(-i)^{N-1}|0(2N-2)f\rangle\big],
\end{aligned}
\end{equation}
where $T_B=\sum^{N-1}_{k=1}t_k'+(N-1)\tau'$. At this moment, part-$B$ is completed. The order of part-$A$ and part-$B$ can be mutually exchanged.

The final step is started by simultaneously tuning the level splittings $\omega_{eg}$ and $\omega_{fg}$ to be resonant with $|(2N-2)0e\rangle \leftrightarrow |(2N)0g\rangle$ and $|0(2N-2)f\rangle \leftrightarrow |0(2N)g\rangle$, respectively. Namely, $\omega_{eg}=2\omega_a+\delta_{2N-2}$ and $\omega_{fg}=2\omega_b+\Delta_{2N-2}$, and now two effective Rabi oscillations are simultaneously switched on. Under the coupling strength $g_{eff}=g'_{eff}$ and the operation time $t_N=\pi/|2g_{eff}(2N-2)|$, the state~\eqref{phitB} eventually becomes
\begin{equation}\label{phitn}
\begin{aligned}
\Psi(T)&=\frac{1}{\sqrt{2}}\left[(-i)^{N}|(2N)0g\rangle+(-i)^{N}|0(2N)g\rangle\right]\\
&=\frac{(-i)^N}{\sqrt{2}}[|(2N)0\rangle+|0(2N)\rangle]\otimes|g\rangle.
\end{aligned}
\end{equation}
The total time for completing our protocol using two-photon resonance is found to be
\begin{equation}\label{T}
\begin{aligned}
T&=T_A+T_B+t_N\\
&=\sum^{N-1}_{j=1}\frac{\pi}{|g_{eff}(2j-2)|}+(N-1)\left(\frac{\pi}{2\Omega_{fg}}+\frac{\pi}{2\Omega_{eg}}\right)
\\ &+\frac{\pi}{2|g_{eff}(2N-2)|}.
\end{aligned}
\end{equation}
Here the time for all the frequency adjustments has been omitted. When $N=1$, the whole procedure of preparation is reduced to the final step, which is exactly the same as illustrated in Sec.~(\ref{twophoton}). Note $\delta_0=\delta$ and $\Delta_0=\Delta$.

The NOON state~\eqref{phitn} can be also constructed when $g_{eff}\neq g_{eff}'$. In this case, we first arrive at $|\Psi(T_A)\rangle$~(\ref{phitA}). Then we perform an extra Step-$A_N$, whose definition is analog to Step-$A_j$ ($j<N$), so that $|\Psi(T_A)\rangle$ evolves to
\begin{equation}\label{psita}
\Psi(\tilde{T}_A)=\frac{1}{\sqrt{2}}\left[(-i)^{N}|(2N)0e\rangle+|00f\rangle\right],
\end{equation}
where $\tilde{T}_A=\sum^{N}_{j=1}t_j+N\tau$. Then we continue the operations in part-$B$. After that, the state~\eqref{psita} becomes
\begin{equation}\label{psitb}
\Psi(\tilde{T}_A+T_B)=\frac{(-i)^{N-1}}{\sqrt{2}}\left[-i|(2N)0e\rangle+|0(2N-2)f\rangle\right].
\end{equation}
Next a pulse of $\{\omega_{eg},\pi/2,\pi/(2\Omega_{eg})\}$ is used to realize $|e\rangle\to|g\rangle$. Finally, the level spacing $\omega_{fg}$ is tuned to be resonant with $|0(2N-2)f\rangle \leftrightarrow |0(2N)g\rangle$. After an evolution time $t_N=\pi/|2g_{eff}'(2N-2)|$, we obtain
\begin{equation}\label{psit}
\Psi(T)=\frac{(-i)^N}{\sqrt{2}}\left[|(2N)0\rangle+|0(2N)\rangle\right]\otimes|g\rangle.
\end{equation}
The time of the whole procedure in total becomes
\begin{equation}\label{Ts}
\begin{aligned}
T&=\sum^{N}_{j=1}\left[\frac{\pi}{2|g_{eff}(2j-2)|}+\frac{\pi}{2|g_{eff}'(2j-2)|}\right]+\frac{N\pi}{2\Omega_{fg}}\\
&+\frac{(N-1)\pi}{2\Omega_{eg}}.
\end{aligned}
\end{equation}

To prepare the NOON states with $2N-1$ (odd number, $N\geq1$) photons, one needs to revise merely the final step after the system arrives at the state in Eq.~\eqref{phitB}. At that moment, we set $\omega_{eg}=\omega_a$, $\omega_{fg}=\omega_b$ and $g_a=g_b$. Then two mutually-independent Rabi oscillations can simultaneously occur in their respective subspaces by virtue of the first-order/single-photon process determined by the interaction Hamiltonian~\eqref{V}. In particular, after the evolution time $t_N=\pi/[2\sqrt{2N-1}g_a\cos\theta]$, the state~\eqref{phitB} becomes
\begin{equation}\label{phitns}
\Psi(T)=\frac{(-i)^N}{\sqrt{2}}\left[|(2N-1)0\rangle+|0(2N-1)\rangle\right]\otimes|g\rangle,
\end{equation}
where
\begin{equation}\label{T2}
\begin{aligned}
T&=T_A+T_B+t_N\\
&=\sum^{N-1}_{j=1}\left[\frac{\pi}{2|g_{eff}(2j-2)|}+\frac{\pi}{2|g_{eff}'(2j-2)|}\right]+(N-1)\\
&\times\left(\frac{\pi}{2\Omega_{fg}}+\frac{\pi}{2\Omega_{eg}}\right)+\frac{\pi}{2\sqrt{2N-1}g_a\cos\theta}.
\end{aligned}
\end{equation}
The results~\eqref{phitn} and~\eqref{phitns} show that the two resonators $a$ and $b$ are eventually prepared in a NOON state.

\begin{figure}[htbp]
\centering
\includegraphics[width=0.4\textwidth]{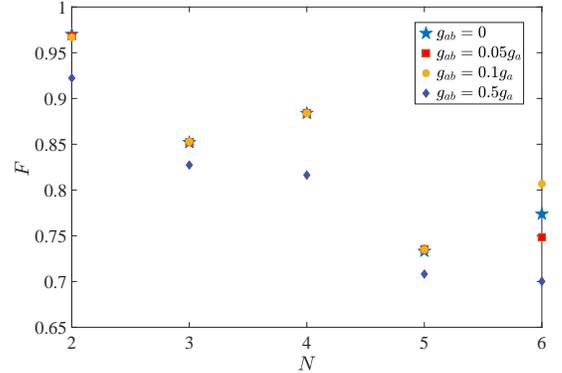}
\caption{(Color online) The fidelity of the NOON state $(|0N\rangle+|N0\rangle)/\sqrt{2}$ under different inter-cavity coupling strengths $g_{ab}$. When $g_{ab}=0.1g_a$, the fidelities are $96.7\%$ ($N=2$), $88.4\%$ ($N=4$), $80.1\%$ ($N=6$) for an even $N$; and $85.3\%$ ($N=3$), $73.5\%$ ($N=5$) for an odd $N$, respectively.}\label{Nfidelity}
\end{figure}

During the state preparation, the whole system is not only subject to the external decoherence channels for each constituent described by the master equation~\eqref{master}, but also under the influence of intrinsic disturbance, such as the crosstalk between the two resonators $a$ and $b$~\cite{crosstalk}. Namely, the interaction Hamiltonian $V$ in Eq.~\eqref{V} can be generalized to
\begin{equation}\label{Vs}
\tilde{V}=V+g_{ab}(a+a^{\dag})(b+b^{\dag}),
\end{equation}
where $g_{ab}$ is the inter-cavity coupling strength. And the full Hamiltonian in Eq.~\eqref{master} is modified accordingly.

In a typical system consisting of a flux qutrit and two resonators~\cite{fq,fq2,fq3,fq4,ca}, the transition frequencies among the three levels of the qutrit can be manipulated within the range $\sim[1, 20]$ GHz. In the numerical evaluation of the fidelity under both decoherence and crosstalk, the frequencies of the resonators $a$ and $b$ are set as $\omega_a/(2\pi)\sim4$GHz and $\omega_b/(2\pi)\sim6.8$GHz, respectively. The decay rates are set as $\gamma_{e}^{-1}=5\mu$s, $\gamma_{f}^{-1}=5\mu$s, $\gamma_{eg}^{-1}=10\mu$s, $\gamma_{fg}^{-1}=20\mu$s, $\gamma_{fe}^{-1}=5\mu$s, and $\kappa_a^{-1}=\kappa_b^{-1}=20\mu$s. The Rabi frequencies of the microwave pulses are set as $\Omega_{eg}=\Omega_{fg}=300$MHz. The effect of the inter-cavity coupling on the NOON-state fidelity is demonstrated in Fig.~\ref{Nfidelity} for different $N$.

We fix the coupling strength of resonator $a$ and the qutrit to be $g_a/(2\pi)=120$MHz, which is about $0.03\omega_a$ according to the previous setting. Note this value is both feasible in experiments and valid for the obtained effective Hamiltonian as shown in Fig.~\ref{two}. Due to the fact that the preparation protocol is parity-dependent, so that for an even $N$, $g_b$ is found to be $\sim69.4$MHz to fulfill the condition $g_{eff}=g'_{eff}$ and for an odd $N$, $g_b=g_a=120$MHz.

\begin{table}
\begin{tabular}{|c|c|c|c|c|c|}
  \hline
  $N$ & 2 & 3 & 4 & 5 & 6 \\
  \hline
  $F(\delta_n, \Delta_n)$ & 92.2\% & 82.7\% & 81.6\% & 70.8\% & 70.0\% \\
  \hline
  $F(\tilde{\delta}_n, \tilde{\Delta}_n)$ & 96.5\% & 84.2\% & 86.6\% & 72.8\% & 75.7\% \\
  \hline
\end{tabular}
\caption{The fidelities of the NOON state $(|0N\rangle+|N0\rangle)/\sqrt{2}$ under $g_{ab}=0.5g_a$ using the energy shifts $\delta_n$ and $\Delta_n$ in Eqs.~\eqref{den} and~\eqref{deln}, respectively or the modified $\tilde{\delta}_n$ and $\tilde{\Delta}_n$ in Eq.~\eqref{deltaab}.}
\label{ftable}
\end{table}

As expected, the fidelities in Fig.~\ref{Nfidelity} decline with the increasing $N$ for either group of parity, i.e., $F(N=2)>F(N=4)>F(N=6)$ and $F(N=3)>F(N=5)$. It turns out that the effect of the crosstalk of cavities could be negligible as long as $g_{ab}\leq0.1g_a$ (it is feasible in experiments~\cite{gab}), by which a high fidelity $\geq 73\%$ can be still obtained for $N\leq5$. When $N=6$, it is interesting to find that $F(g_{ab}=0.1g_a)>F(g_{ab}=0)>F(g_{ab}=0.05g_a)>F(g_{ab}=0.5g_a)$, which indicates the deficiency of the preceding effective Hamiltonian. To accommodate a larger $N$ and a stronger inter-cavity coupling, we could revisit the perturbative treatment in Appendix~\ref{appa} using the updated full Hamiltonian with the interaction Hamiltonian~(\ref{Vs}). It gives rises to the effective Hamiltonians in the same formation as in Eqs.~(\ref{Heff}) and (\ref{Heffs}), but with modified energy shifts:
\begin{equation}\label{deltaab}
\begin{aligned}
\tilde{\delta}_n=\delta_n+2g^2_{ab}\left(\frac{1}{\omega_a-\omega_b}-\frac{1}{\omega_a+\omega_b}\right), \\
\tilde{\Delta}_n=\Delta_n+2g^2_{ab}\left(\frac{1}{\omega_b-\omega_a}-\frac{1}{\omega_a+\omega_b}\right),
\end{aligned}
\end{equation}
which have taken account the effect of inter-cavity crosstalk into the double-photon resonant points of $\omega_{eg}$ and $\omega_{fg}$, respectively. As demonstrated in Table~\ref{ftable}, the application of the modified $\tilde{\delta}_n$ and $\tilde{\Delta}_n$ can really improve the fidelities with different $N$ even under $g_{ab}=0.5g_a$.

\section{Discussion and Conclusion}\label{dc}

The effective Hamiltonians obtained in Sec.~\ref{model} respectively in charge of the two-photon resonance (see Sec.~\ref{twophoton}) and the three-photon resonance (see Sec.~\ref{threephoton}) capture the effects from the relevant transitions in the whole Hilbert space and have been justified by the numerical evaluation. The energy shifts in the leading order $\delta$, $\Delta$, $\delta_s$ and $\Delta_s$ respectively presented in Eqs.~(\ref{de}), (\ref{De}), (\ref{delta}) and (\ref{Delta}) determine the points for the intrinsic multi-photon resonances. Subsequently the effective coupling strengthes $g_{eff}$, $g_{eff}'$, $g_{effs}$ and $g_{effs}'$ respectively presented in Eqs.~(\ref{ge}), (\ref{ges}), (\ref{eg}) and (\ref{egs}) are used to evaluate the period of the desired Rabi oscillations.

More than the protocols for constructing the NOON states with $N=2$ and $N=3$, the effective Hamiltonians are the basic elements in generating the NOON states with more photons in Sec.~\ref{pp}. Comparing to the previous protocol~\cite{noon4} to the NOON state with the same number of photons, ours in Sec.~\ref{pp} reduces the number of operations to about a half. The state fidelity under preparation has been estimated by a Lindblad master equation in the eigenbases of the original Hamiltonian of Sec.~\ref{fide}.

The protocol we proposed is essentially based on two independent and parallel Rabi oscillations in the $\triangle$-type qutrit, which can adapt to the other types of three-level systems, such as the $V$-type and the ladder-type qutrit~\cite{noon4}, via a properly modified interaction Hamiltonian $V$. For example, the interaction Hamiltonian for the $V$-type three-level system has no terms containing $\sigma^{fe}_{x,z}$. While a more complex $\triangle$-type qutrit model established in the circuit-QED system is a compromise with respect to the practical realization by the state-of-art quantum technology. The first reason for this choice is that the excited states of the atomic system are able to be adjusted on demand, which greatly reduces the possibility of choosing the natural atoms. The second one is that to realize the target NOON state as fast as possible (otherwise the accumulated decoherence effect will destroy the target state), the coupling strength between the atomic system and the resonators should be of the strong regime. The circuit-QED system satisfies these two conditions; and in many situations the three-level system of the superconduting circuit is of a $\triangle$-type qutrit.

In conclusion, we have presented a concise protocol for the deterministic preparation of the NOON states. We do this in a setup consisting of two resonator modes strongly coupled to a single $\triangle$-type superconducting qutrit within the framework of a general Rabi model. The protocol relies on the effective Hamiltonians at the avoided level-crossing points, which reserves the effects of the counter-rotating terms and the leading-order contributions of the state transitions. By properly using the effective Hamiltonian, the population at the excited states of the qutrit pumped by the microwave driving pulses can be transferred to multiple photons in the corresponding resonator modes. Moreover, our protocol is found to be robust against the external decoherence with a typical magnitude of decay rate and the internal crosstalk of the resonators. Hence, our study is of interested in pursuit of the entangled state with the counter-rotating interaction and of important to control the quantum state in the circuit-QED system with fewer steps and fewer devices.

\section*{Acknowledgments}

We acknowledge grant support from the National Science Foundation of China (Grants No. 11974311 and No. U1801661), and Zhejiang Provincial Natural Science Foundation of China under Grant No. LD18A040001.

\appendix

\section{The effective Hamiltonian for the two-photon resonances} \label{appa}

\begin{figure}[htbp]
\centering
\includegraphics[width=0.4\textwidth]{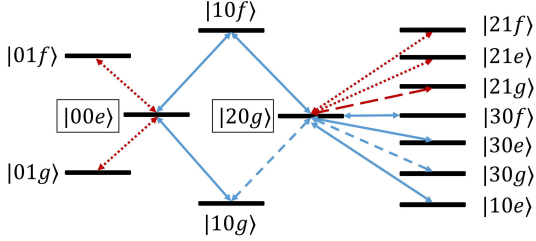}
\caption{(Color online) All the second-order (leading-order) paths involving the bases $|00e\rangle$ and $|20g\rangle$. Blue solid (dashed) lines mark the transitions mediated by $\sigma_x$ ($\sigma_z$) relative to the resonator $a$. Red dotted (long-dashed) lines mark the transitions mediated by $\sigma_x$ ($\sigma_z$) relative to the resonator $b$. }\label{twopath}
\end{figure}

To extract the effective Hamiltonian~\eqref{Heff} in the subspace spanned by $\{|00e\rangle\equiv|0\rangle_a|0\rangle_b|e\rangle, |20g\rangle\equiv|2\rangle_a|0\rangle_b|g\rangle\}$ from the full Hamiltonian~(\ref{noonmodel}) for the two-photon resonance shown in Fig.~\ref{two}(a), one can apply the standard perturbation theory with respect to the atom-photon coupling strengths $g_a$ and $g_b$. In Fig.~\ref{twopath}, all the second-order processes involving with these two bases will be considered in the following construction of the effective Hamiltonian.

The interaction Hamiltonian $V$ in Eq.~\eqref{V} can be regarded as a perturbation provided that $g_a, g_b\ll\omega_a, \omega_b, |\omega_b-\omega_a|$. To the second order, the effective coupling strength or energy shift between any eigenstates $|i\rangle$ and $|j\rangle$ of the unperturbed Hamiltonian $H_0$ in Eq.~\eqref{noonmodel} is given by~\cite{sc2,bg}
\begin{equation}\label{secondp}
g=\sum_{n\neq i,j}\frac{V_{jn}V_{ni}}{\omega_i-\omega_n}
\end{equation}
where $V_{nm}\equiv\langle n|V|m\rangle$ and $\omega_n$ is the eigenenergy of $|n\rangle$.

We write the avoid level-crossing point $\omega_{eg}=2\omega_a+\delta$, where $\delta$ is an undetermined energy shift consistent with the second-order effective Hamiltonian. It is worthy to emphasis this shift should not be omitted as in some literatures, since it is in the same order of the effective coupling strength. We first consider the contribution to the effective coupling strength from the two paths connecting $|00e\rangle$ and $|20g\rangle$, i.e., $|00e\rangle \to |10f\rangle \to |20g\rangle$ and $|00e\rangle \to |10g\rangle \to |20g\rangle$ as shown in Fig.~\ref{twopath}. By virtue of Eq.~(\ref{secondp}), one can get
\begin{equation}\label{g12}
\begin{aligned}
g_{eff}&=-\sqrt{2}g^2_a\left[\frac{\sin(2\theta)}{\omega_a+\delta}+\frac{\cos^2\theta}{\omega_{fg}-\omega_a-\delta}\right]\\
&=-\sqrt{2}g^2_a\left[\frac{\sin(2\theta)}{\omega_a}+\frac{\cos^2\theta}{\omega_{fg}-\omega_a}\right]+\mathcal{O}(\delta),
\end{aligned}
\end{equation}
where $\mathcal{O}(\delta)$ means all the other orders of $\delta$ from the zeroth order (the first term of the second line) in terms of Taylor expansion. The other paths in Fig.~\ref{twopath} (For example, $|00e\rangle \to |01f\rangle \to |00e\rangle$) are in charge of the energy shifts for $|00e\rangle$ or $|20g\rangle$.

Summarizing all the four paths from $|00e\rangle$ and back to $|00e\rangle$ through a mediate state, i.e., $|00e\rangle\to|10g\rangle\to|00e\rangle$, $|00e\rangle\to|10f\rangle\to|00e\rangle$, $|00e\rangle\to|01g\rangle\to|00e\rangle$, and $|00e\rangle\to|01f\rangle\to|00e\rangle$, one can obtain the second-order energy correction (shift) $\epsilon_1$ for the state $|00e\rangle$ according to Eq.~\eqref{secondp}
\begin{equation}\label{g11}
\begin{aligned}
\epsilon_1&=g^2_a\cos^2\theta\left(\frac{1}{\omega_a}-\frac{1}{\omega_{fg}-\omega_a}\right)\\
&+g^2_b\cos^2\theta\left(\frac{1}{2\omega_a-\omega_b}-\frac{1}{\omega_{fg}+\omega_b-2\omega_a}\right)+\mathcal{O}(\delta).
\end{aligned}
\end{equation}

And in the same way, the energy shift $\epsilon_2$ for the state $|20g\rangle$ is found to be
\begin{equation}\label{g22}
\begin{aligned}
\epsilon_2&=-g^2_a\cos^2\theta\left(\frac{3}{\omega_a}+\frac{2}{\omega_{fg}-\omega_a}+\frac{3}{\omega_{fg}+\omega_a}\right)\\
&-g^2_b\cos^2\theta\left(\frac{1}{2\omega_a+\omega_b}+\frac{1}{\omega_{fg}+\omega_b}\right)-\frac{4g^2_a\sin^2\theta}{\omega_a}\\
&-\frac{4g^2_b\sin^2\theta}{\omega_b}+\mathcal{O}(\delta).
\end{aligned}
\end{equation}

The collection of the results in Eqs.~\eqref{g12}, \eqref{g11}, and \eqref{g22} gives rise to the second-order effective Hamiltonian:
\begin{equation}\label{He12}
\begin{aligned}
H_{eff}&=(\omega_{eg}+\epsilon_1)|00e\rangle\langle 00e|+(2\omega_a+\epsilon_2)|20g\rangle\langle20g| \\
&+g_{eff}(|00e\rangle\langle 20g|+|20g\rangle\langle 00e|).
\end{aligned}
\end{equation}
An exact double-resonance facilitated by Eq.~(\ref{He12}) allows a completed Rabi oscillation between $|00e\rangle$ and $|20g\rangle$, which requires that the first line of Eq.~(\ref{He12}) becomes an identity operator in the very subspace. Thus $\omega_{eg}+\epsilon_1=2\omega_a+\epsilon_2$. Recalling the assumption that $\omega_{eg}=2\omega_a+\delta$, one can then obtain
\begin{equation}\label{deltatwo}
\begin{aligned}
\delta&=\epsilon_2-\epsilon_1\\
&=-g^2_a\cos^2\theta\left(\frac{4}{\omega_a}+\frac{1}{\omega_{fg}-\omega_a}+\frac{3}{\omega_{fg}+\omega_a}\right) \\
&-g^2_b\cos^2\theta\bigg(\frac{1}{2\omega_a+\omega_b}+\frac{1}{\omega_{fg}+\omega_b}+\frac{1}{2\omega_a-\omega_b} \\
&-\frac{1}{\omega_{fg}+\omega_b-2\omega_a}\bigg)-\frac{4g^2_a\sin^2\theta}{\omega_a}-\frac{4g^2_b\sin^2\theta}{\omega_b}.
\end{aligned}
\end{equation}
Eventually the effective Hamiltonian~\eqref{He12} can be written as
\begin{equation}\label{Heff12}
H_{eff}=g_{eff}(|00e\rangle\langle 20g|+|20g\rangle\langle 00e|),
\end{equation}
which is Eq.~\eqref{Heff} in the main text. The foregoing perturbative calculation prohibits $\omega_b\approx k\omega_a$ and $\omega_{fg}\approx k\omega_a$ with $k$ integer. Otherwise, the effect from the other degenerate states can not be omitted. For example, when $\omega_a\approx\omega_b$, the states $|20g\rangle$ and $|02g\rangle$ can not be differentiated by the qutrit.

\begin{figure}[htbp]
\centering
\includegraphics[width=0.4\textwidth]{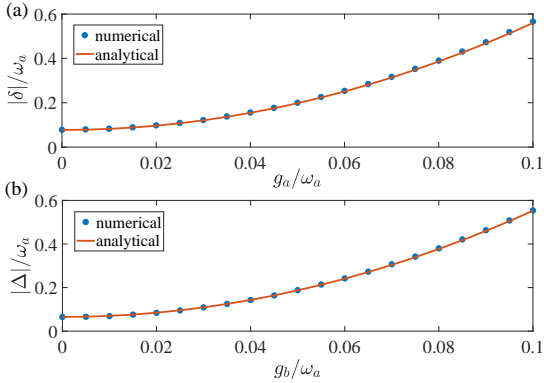}
\caption{(Color online) (a) and (b) Comparison between the numerically evaluated normalized energy shifts from the full Hamiltonian (blue points) and the corresponding analytical results from the second-order effective Hamiltonian (red solid line) in Eqs.~\eqref{deltatwo} and \eqref{Deltatwo}, respectively. Here $\omega_b=1.7\omega_a$ and $\theta=\frac{\pi}{6}$. For (a) $g_b=0.05\omega_a$, $\omega_{fg}=2\omega_b$; for (b) $g_a=0.05\omega_a$, $\omega_{eg}=2\omega_a$.}\label{twodelta}
\end{figure}

A similar derivation yields the effective Hamiltonian~\eqref{Heffs} for the two-photon resonance shown in Fig.~\ref{two}(c), which lives in the subspace spanned by $\{|00f\rangle,|02g\rangle\}$. Again the avoid level-crossing point $\omega_{fg}$ is expressed by $\omega_{fg}=2\omega_b+\Delta$ with $|\Delta|\ll \omega_a, \omega_b, |\omega_b-\omega_a|$ the undetermined energy shift. The effective Hamiltonian can be written as
\begin{equation}\label{He34}
\begin{aligned}
H_{eff}'&=(\omega_{fg}+\epsilon_3)|00f\rangle\langle 00f|+(2\omega_b+\epsilon_4)|02g\rangle\langle02g| \\
&+g_{eff}'(|00f\rangle\langle 02g|+|02g\rangle\langle 00f|),
\end{aligned}
\end{equation}
where the effective coupling strength connecting $|00f\rangle$ and $|02g\rangle$ is
\begin{equation}\label{g34}
g_{eff}'=-\sqrt{2}g^2_b\left[\frac{2\sin(2\theta)}{\omega_b}+\frac{\cos^2\theta}{\omega_{eg}-\omega_{b}}\right]
+\mathcal{O}(\Delta),
\end{equation}
the energy shift for the state $|00f\rangle$ is
\begin{equation}\label{g33}
\begin{aligned}
\epsilon_3&= g^2_a\cos^2\theta\left(\frac{1}{2\omega_b-\omega_a-\omega_{eg}}+\frac{1}{2\omega_b-\omega_a}\right)\\
&+g^2_b\cos^2\theta\left(\frac{1}{\omega_b-\omega_{eg}}+\frac{1}{\omega_b}\right)-\frac{4g^2_a\sin^2\theta}{\omega_a} \\
&-\frac{4g^2_b\sin^2\theta}{\omega_b}+\mathcal{O}(\Delta),
\end{aligned}
\end{equation}
and the energy shift for the state $|02g\rangle$ is
\begin{equation}\label{g44}
\begin{aligned}
\epsilon_4&=-g^2_a\cos^2\theta\left(\frac{1}{\omega_{eg}+\omega_a}+\frac{1}{2\omega_b+\omega_a}\right)\\
&-g^2_b\cos^2\theta\left(\frac{3}{\omega_b}+\frac{2}{\omega_{eg}-\omega_b}+\frac{3}{\omega_{eg}+\omega_b}\right)\\
&-\frac{4g^2_a\sin^2\theta}{\omega_a}-\frac{4g^2_b\sin^2\theta}{\omega_b}+\mathcal{O}(\Delta).
\end{aligned}
\end{equation}

Again, to realize a completed Rabi oscillation between $|00f\rangle$ and $|02g\rangle$, $\omega_{fg}+\epsilon_3$ should be equivalent to $2\omega_b+\epsilon_4$. It turns out that
\begin{equation}\label{Deltatwo}
\begin{aligned}
\Delta&=\epsilon_4-\epsilon_3\\
&=-g^2_a\cos^2\theta\bigg(\frac{1}{\omega_{eg}+\omega_a}+\frac{1}{2\omega_b+\omega_a}+\frac{1}{2\omega_b-\omega_a}\\
&+\frac{1}{2\omega_b-\omega_a-\omega_{eg}}\bigg)-g^2_b\cos^2\theta\Big(\frac{4}{\omega_b}+\frac{1}{\omega_{eg}-\omega_b}\\
&+\frac{3}{\omega_{eg}+\omega_b}\Big).
\end{aligned}
\end{equation}
Then the effective Hamiltonian~\eqref{He34} is written as
\begin{equation}\label{Heff34}
H_{eff}'=g_{eff}'\left(|00f\rangle\langle 02g|+|02g\rangle\langle 00f|\right).
\end{equation}

The two energy shifts in Eqs.~\eqref{deltatwo} and \eqref{Deltatwo} can be justified by Fig.~\ref{twodelta}. The normalized $\delta$ ($\Delta$) as a function of $g_a$ ($g_b$) is compared with that from the standard diagonalization of the full Hamiltonian~\eqref{noonmodel}. It is shown that the analytical results do match with the numerical ones at least for normalized atom-photon interaction strength $g_a, g_b/\omega_a \leq 0.1$.

\section{The effective Hamiltonian for the three-photon resonances}\label{appb}

\begin{figure}[htbp]
\centering
\includegraphics[width=0.4\textwidth]{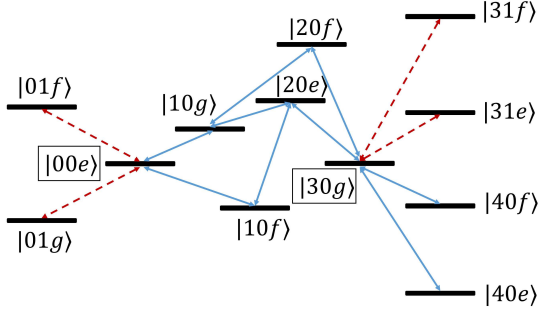}
\caption{(Color online) All the leading-order paths involving the bases $|00e\rangle$ and $|30g\rangle$. Blue solid lines mark the transitions mediated by the coupling to the resonator $a$, which are charge of the connections between $|00e\rangle$ and $|30g\rangle$. Red dashed lines mark the transition mediated by the coupling to the resonator $b$, which are charge of the energy shift for $|00e\rangle$ or $|30g\rangle$.}\label{threepath}
\end{figure}

To construct the effective Hamiltonian~\eqref{effH} from the full Hamiltonian~\eqref{threemodel} in the subspace spanned by $\{|00e\rangle, |30g\rangle\}$ for the three-photon resonance remarked in Fig.~\ref{three}(a). The leading-order paths plotted in Fig.~\ref{threepath} are either the paths of the three-order processes connecting $|00e\rangle$ and $|30g\rangle$ or those of the two-order processes in charge of the energy shifts for these two bases. Due to the standard perturbation theory, the effective coupling strength between any eigenstates $|i\rangle$ and $|j\rangle$ of the unperturbed Hamiltonian $H_0$~\eqref{threemodel} is given by~\cite{sc2,bg}
\begin{equation}\label{thirdp}
g_{s}=\sum_{n,m\neq i,j}\frac{V_{jn}V_{nm}V_{mi}}{(\omega_i-\omega_n)(\omega_i-\omega_m)}.
\end{equation}

The avoid level crossing point can be written as $\omega_{eg}=3\omega_a+\delta_s$, where $\delta_s$ is a to-be-determined energy shift. Considering all the three paths (shown in Fig.~\ref{threepath}) connecting the basis $|00e\rangle$ and the basis $|30g\rangle$, i.e., $|00e\rangle\to|10g\rangle\to|20f\rangle\to|30g\rangle$, $|00e\rangle\to|10g\rangle\to|20e\rangle\to|30g\rangle$ and $|00e\rangle\to|10f\rangle\to|20e\rangle\to|30g\rangle$, we can find the effective coupling strength by using Eq.~\eqref{thirdp},
\begin{equation}\label{effg12}
\begin{aligned}
g_{effs}&=-\sqrt{6}g^3_a\bigg[\frac{1}{2\omega_a(2\omega_a+\delta_s)}-\frac{1}{2\omega_a(\omega_{fg}-2\omega_a-\delta_s)}\\
&+\frac{1}{(2\omega_a+\delta_s)(\omega_{fg}-\omega_a-\delta_s)}\bigg]\\
&=-\frac{\sqrt{6}g^3_a}{2\omega_a}\left(\frac{1}{2\omega_a}
-\frac{1}{\omega_{fg}-2\omega_a}+\frac{1}{\omega_{fg}-\omega_a}\right)\\
&+\mathcal{O}(\delta_s).
\end{aligned}
\end{equation}

Summarizing all the paths from the state $|00e\rangle$ and back to itself through a mediate state, one can obtain its second-order energy shift $\epsilon_{s1}$ according to Eq.~\eqref{secondp}
\begin{equation}\label{effg11}
\begin{aligned}
\epsilon_{s1}&=g^2_a\left(\frac{1}{2\omega_a}+\frac{1}{2\omega_a-\omega_{fg}}\right)\\
&+g^2_b\left(\frac{1}{3\omega_a-\omega_b}+\frac{1}{3\omega_a-\omega_b-\omega_{fg}}\right)+\mathcal{O}(\delta_s).
\end{aligned}
\end{equation}
In the same way, the energy shift $\epsilon_{s2}$ for the basis $|30g\rangle$ is found to be
\begin{equation}\label{effg22}
\begin{aligned}
\epsilon_{s2}&=-g^2_a\left(\frac{5}{2\omega_a}+\frac{3}{\omega_{fg}-\omega_a}+\frac{4}{\omega_{fg}+\omega_a}\right)\\
&-g^2_b\left(\frac{1}{3\omega_a+\omega_b}+\frac{1}{\omega_b+\omega_{fg}}\right)+\mathcal{O}(\delta_s).
\end{aligned}
\end{equation}
With Eqs.~(\ref{effg12}), (\ref{effg11}) and (\ref{effg22}), the effective Hamiltonian in the subspace spanned by $\{|00e\rangle, |30g\rangle\}$ is
\begin{equation}\label{eH12}
\begin{aligned}
H_{effs}&=(\omega_{eg}+\epsilon_{s1})|00e\rangle\langle 00e|+(3\omega_a+\epsilon_{s2})|30g\rangle\langle30g|\\
&+g_{effs}(|00e\rangle\langle 30g|+|30g\rangle\langle 00e|).
\end{aligned}
\end{equation}
Similar to the double-photon resonance treatment in Appendix~\ref{appa}, the first line of Eq.~(\ref{eH12}) should be an effective identity operator in the very subspace to facilitate a completed Rabi oscillation driven by the second line. It is achieved by equating the diagonal elements of~\eqref{eH12}, which gives
\begin{equation}\label{deltathree}
\begin{aligned}
\delta_s&=\epsilon_{s2}-\epsilon_{s1}\\
&=-g^2_a\left(\frac{3}{\omega_a}+\frac{3}{\omega_{fg}-\omega_a}+\frac{4}{\omega_{fg}+\omega_a}
+\frac{1}{2\omega_a-\omega_{fg}}\right)\\
&-g^2_b\bigg(\frac{1}{3\omega_a+\omega_b}+\frac{1}{\omega_b+\omega_{fg}}+\frac{1}{3\omega_a-\omega_b}\\
&+\frac{1}{3\omega_a-\omega_b-\omega_{fg}}\bigg).
\end{aligned}
\end{equation}
Eventually the effective Hamiltonian~\eqref{eH12} becomes
\begin{equation}\label{effH12}
H_{effs}=g_{effs}(|00e\rangle\langle 30g|+|30g\rangle\langle 00e|)
\end{equation}
which is Eq.~\eqref{effH} in the main text.

\begin{figure}[htbp]
\centering
\includegraphics[width=0.4\textwidth]{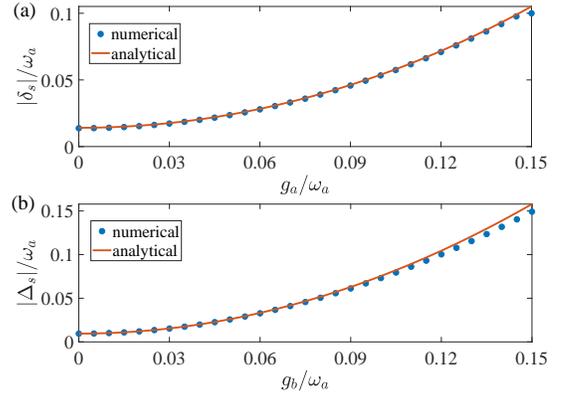}
\caption{(Color online) (a) and (b) Comparison between the numerically calculated normalized energy shifts from the full Hamiltonian (blue points) and the corresponding analytical results from the leading-order effective Hamiltonian (red solid line) in Eqs.~\eqref{deltathree} and \eqref{Deltathree}, respectively. Here $\omega_b=2.2\omega_a$. For (a) $g_b=0.05\omega_a$, $\omega_{fg}=3\omega_b$; for (b) $g_a=0.05\omega_a$, $\omega_{eg}=3\omega_a$.}\label{threedelta}
\end{figure}

In parallel, one can derive the effective Hamiltonian~\eqref{effHs} in the subspace spanned by $\{|00f\rangle, |03g\rangle\}$ at the avoided level crossing point $\omega_{fg}=3\omega_b+\Delta_s$ as shown in Fig.~\ref{three}(c). The effective Hamiltonian is found to be
\begin{equation}\label{effH34}
\begin{aligned}
H_{effs}'&=(\omega_{fg}+\epsilon_{s3})|00f\rangle\langle 00f|+(3\omega_b+\epsilon_{s4})|03g\rangle\langle03g|\\
&+g_{effs}'(|00f\rangle\langle 03g|+|03g\rangle\langle 00f|),
\end{aligned}
\end{equation}
where the effective coupling strength is
\begin{equation}\label{effg34}
\begin{aligned}
g_{effs}'&=-\frac{\sqrt{6}g^3_b}{2\omega_b}\left(\frac{1}{2\omega_b}-\frac{1}{\omega_{eg}-2\omega_b}+\frac{1}{\omega_{eg}-\omega_b}\right)\\
&+\mathcal{O}(\Delta_s),
\end{aligned}
\end{equation}
and the energy shifts for $|00f\rangle$ and $|03g\rangle$ are
\begin{equation}\label{effg33}
\begin{aligned}
\epsilon_{s3}&=g^2_a\left(\frac{1}{3\omega_b-\omega_a-\omega_{eg}}+\frac{1}{3\omega_b-\omega_a}\right)\\
&+g^2_b\left(\frac{1}{2\omega_b-\omega_{eg}}+\frac{1}{2\omega_b}\right)+O(\Delta_s),\\
\epsilon_{s4}&=-g^2_a\left(\frac{1}{3\omega_b+\omega_a}+\frac{1}{\omega_a+\omega_{eg}}\right)\\
&-g^2_b\left(\frac{5}{2\omega_b}+\frac{3}{\omega_{eg}-\omega_b}+\frac{4}{\omega_{eg}+\omega_b}\right)+\mathcal{O}(\Delta_s),
\end{aligned}
\end{equation}
respectively.

Again, to realize a completed Rabi oscillation between $|00f\rangle$ and $|02g\rangle$, we let $\omega_{fg}+\epsilon_3=3\omega_b+\epsilon_4$. It turns out that
\begin{equation}\label{Deltathree}
\begin{aligned}
\Delta_s&=\epsilon_{s4}-\epsilon_{s3}\\
&=-g^2_a\bigg(\frac{1}{3\omega_b+\omega_a}+\frac{1}{\omega_a+\omega_{eg}}+\frac{1}{3\omega_b-\omega_a-\omega_{eg}}\\
&+\frac{1}{3\omega_b-\omega_a}\bigg)-g^2_b\bigg(\frac{1}{2\omega_b-\omega_{eg}}+\frac{3}{\omega_b}+\frac{3}{\omega_{eg}-\omega_b}\\
&+\frac{4}{\omega_{eg}+\omega_b}\bigg).
\end{aligned}
\end{equation}
Then we get the effective Hamiltonian~\eqref{effHs} in the main text:
\begin{equation}\label{eH34}
H_{effs}'=g_{effs}'(|00f\rangle\langle 03g|+|03g\rangle\langle 00f|).
\end{equation}

In Fig.~\ref{threedelta}, we demonstrate the analytical results in Eqs.~\eqref{deltathree} and~\eqref{Deltathree}. The normalized $\delta_s$ ($\Delta_s$) as a function of $g_a$ ($g_b$) is compared to that from the standard diagonalization of the full Hamiltonian in Eq.~\eqref{threemodel}. It is shown that the analytical results do match with the numerical ones at least for normalized atom-photon interaction strength $g_a/\omega_a\leq0.14$ and $g_b/\omega_a\leq0.11$. They are in the strong-coupling regime.

\bibliographystyle{apsrevlong}
\bibliography{references}

\begin{thebibliography}{49}%
\makeatletter
\providecommand \@ifxundefined [1]{%
 \@ifx{#1\undefined}
}%
\providecommand \@ifnum [1]{%
 \ifnum #1\expandafter \@firstoftwo
 \else \expandafter \@secondoftwo
 \fi
}%
\providecommand \@ifx [1]{%
 \ifx #1\expandafter \@firstoftwo
 \else \expandafter \@secondoftwo
 \fi
}%
\providecommand \natexlab [1]{#1}%
\providecommand \enquote  [1]{``#1''}%
\providecommand \bibnamefont  [1]{#1}%
\providecommand \bibfnamefont [1]{#1}%
\providecommand \citenamefont [1]{#1}%
\providecommand \href@noop [0]{\@secondoftwo}%
\providecommand \href [0]{\begingroup \@sanitize@url \@href}%
\providecommand \@href[1]{\@@startlink{#1}\@@href}%
\providecommand \@@href[1]{\endgroup#1\@@endlink}%
\providecommand \@sanitize@url [0]{\catcode `\\12\catcode `\$12\catcode
  `\&12\catcode `\#12\catcode `\^12\catcode `\_12\catcode `\%12\relax}%
\providecommand \@@startlink[1]{}%
\providecommand \@@endlink[0]{}%
\providecommand \url  [0]{\begingroup\@sanitize@url \@url }%
\providecommand \@url [1]{\endgroup\@href {#1}{\urlprefix }}%
\providecommand \urlprefix  [0]{URL }%
\providecommand \Eprint [0]{\href }%
\providecommand \doibase [0]{http://dx.doi.org/}%
\providecommand \selectlanguage [0]{\@gobble}%
\providecommand \bibinfo  [0]{\@secondoftwo}%
\providecommand \bibfield  [0]{\@secondoftwo}%
\providecommand \translation [1]{[#1]}%
\providecommand \BibitemOpen [0]{}%
\providecommand \bibitemStop [0]{}%
\providecommand \bibitemNoStop [0]{.\EOS\space}%
\providecommand \EOS [0]{\spacefactor3000\relax}%
\providecommand \BibitemShut  [1]{\csname bibitem#1\endcsname}%
\let\auto@bib@innerbib\@empty
\bibitem [{\citenamefont {Horodecki}\ \emph {et~al.}(2009)\citenamefont
  {Horodecki}, \citenamefont {Horodecki}, \citenamefont {Horodecki},\ and\
  \citenamefont {Horodecki}}]{qe}%
  \BibitemOpen
  \bibfield  {author} {\bibinfo {author} {\bibfnamefont {R.}~\bibnamefont
  {Horodecki}}, \bibinfo {author} {\bibfnamefont {P.}~\bibnamefont
  {Horodecki}}, \bibinfo {author} {\bibfnamefont {M.}~\bibnamefont
  {Horodecki}}, \ and\ \bibinfo {author} {\bibfnamefont {K.}~\bibnamefont
  {Horodecki}},\ }\bibfield  {title} {\emph {\bibinfo {title} {Quantum
  entanglement},\ }}\href {\doibase 10.1103/RevModPhys.81.865} {\bibfield
  {journal} {\bibinfo  {journal} {Rev. Mod. Phys.}\ }\textbf {\bibinfo {volume}
  {81}},\ \bibinfo {pages} {865} (\bibinfo {year} {2009})}\BibitemShut
  {NoStop}%
\bibitem [{\citenamefont {Gisin}\ and\ \citenamefont {Thew}(2007)}]{qco}%
  \BibitemOpen
  \bibfield  {author} {\bibinfo {author} {\bibfnamefont {N.}~\bibnamefont
  {Gisin}}\ and\ \bibinfo {author} {\bibfnamefont {R.}~\bibnamefont {Thew}},\
  }\bibfield  {title} {\emph {\bibinfo {title} {Quantum communication},\
  }}\href@noop {} {\bibfield  {journal} {\bibinfo  {journal} {Nat. Photon.}\
  }\textbf {\bibinfo {volume} {1}},\ \bibinfo {pages} {165} (\bibinfo {year}
  {2007})}\BibitemShut {NoStop}%
\bibitem [{\citenamefont {Xiu}\ \emph {et~al.}(2017)\citenamefont {Xiu},
  \citenamefont {Kockum}, \citenamefont {Miranowicz}, \citenamefont {Liu},\
  and\ \citenamefont {Nori}}]{in}%
  \BibitemOpen
  \bibfield  {author} {\bibinfo {author} {\bibfnamefont {G.}~\bibnamefont
  {Xiu}}, \bibinfo {author} {\bibfnamefont {A.~F.}\ \bibnamefont {Kockum}},
  \bibinfo {author} {\bibfnamefont {A.}~\bibnamefont {Miranowicz}}, \bibinfo
  {author} {\bibfnamefont {Y.~X.}\ \bibnamefont {Liu}}, \ and\ \bibinfo
  {author} {\bibfnamefont {F.}~\bibnamefont {Nori}},\ }\bibfield  {title}
  {\emph {\bibinfo {title} {Microwave photonics with superconducting quantum
  circuits},\ }}\href@noop {} {\bibfield  {journal} {\bibinfo  {journal} {Phys.
  Rep.}\ }\textbf {\bibinfo {volume} {718}},\ \bibinfo {pages} {1} (\bibinfo
  {year} {2017})}\BibitemShut {NoStop}%
\bibitem [{\citenamefont {Ekert}(1991)}]{kd}%
  \BibitemOpen
  \bibfield  {author} {\bibinfo {author} {\bibfnamefont {A.~K.}\ \bibnamefont
  {Ekert}},\ }\bibfield  {title} {\emph {\bibinfo {title} {Quantum cryptography
  based on bell's theorem},\ }}\href {\doibase 10.1103/PhysRevLett.67.661}
  {\bibfield  {journal} {\bibinfo  {journal} {Phys. Rev. Lett.}\ }\textbf
  {\bibinfo {volume} {67}},\ \bibinfo {pages} {661} (\bibinfo {year}
  {1991})}\BibitemShut {NoStop}%
\bibitem [{\citenamefont {Hillery}\ \emph {et~al.}(1999)\citenamefont
  {Hillery}, \citenamefont {Bu\ifmmode~\check{z}\else \v{z}\fi{}ek},\ and\
  \citenamefont {Berthiaume}}]{qss}%
  \BibitemOpen
  \bibfield  {author} {\bibinfo {author} {\bibfnamefont {M.}~\bibnamefont
  {Hillery}}, \bibinfo {author} {\bibfnamefont {V.}~\bibnamefont
  {Bu\ifmmode~\check{z}\else \v{z}\fi{}ek}}, \ and\ \bibinfo {author}
  {\bibfnamefont {A.}~\bibnamefont {Berthiaume}},\ }\bibfield  {title} {\emph
  {\bibinfo {title} {Quantum secret sharing},\ }}\href {\doibase
  10.1103/PhysRevA.59.1829} {\bibfield  {journal} {\bibinfo  {journal} {Phys.
  Rev. A}\ }\textbf {\bibinfo {volume} {59}},\ \bibinfo {pages} {1829}
  (\bibinfo {year} {1999})}\BibitemShut {NoStop}%
\bibitem [{\citenamefont {Long}\ and\ \citenamefont {Liu}(2002)}]{qsc}%
  \BibitemOpen
  \bibfield  {author} {\bibinfo {author} {\bibfnamefont {G.~L.}\ \bibnamefont
  {Long}}\ and\ \bibinfo {author} {\bibfnamefont {X.~S.}\ \bibnamefont {Liu}},\
  }\bibfield  {title} {\emph {\bibinfo {title} {Theoretically efficient
  high-capacity quantum-key-distribution scheme},\ }}\href {\doibase
  10.1103/PhysRevA.65.032302} {\bibfield  {journal} {\bibinfo  {journal} {Phys.
  Rev. A}\ }\textbf {\bibinfo {volume} {65}},\ \bibinfo {pages} {032302}
  (\bibinfo {year} {2002})}\BibitemShut {NoStop}%
\bibitem [{\citenamefont {Deng}\ \emph {et~al.}(2003)\citenamefont {Deng},
  \citenamefont {Long},\ and\ \citenamefont {Liu}}]{qsc2}%
  \BibitemOpen
  \bibfield  {author} {\bibinfo {author} {\bibfnamefont {F.~G.}\ \bibnamefont
  {Deng}}, \bibinfo {author} {\bibfnamefont {G.~L.}\ \bibnamefont {Long}}, \
  and\ \bibinfo {author} {\bibfnamefont {X.-S.}\ \bibnamefont {Liu}},\
  }\bibfield  {title} {\emph {\bibinfo {title} {Two-step quantum direct
  communication protocol using the einstein-podolsky-rosen pair block},\
  }}\href {\doibase 10.1103/PhysRevA.68.042317} {\bibfield  {journal} {\bibinfo
   {journal} {Phys. Rev. A}\ }\textbf {\bibinfo {volume} {68}},\ \bibinfo
  {pages} {042317} (\bibinfo {year} {2003})}\BibitemShut {NoStop}%
\bibitem [{\citenamefont {Wei}\ \emph {et~al.}(2006)\citenamefont {Wei},
  \citenamefont {Liu},\ and\ \citenamefont {Nori}}]{ghz}%
  \BibitemOpen
  \bibfield  {author} {\bibinfo {author} {\bibfnamefont {L.~F.}\ \bibnamefont
  {Wei}}, \bibinfo {author} {\bibfnamefont {Y.-x.}\ \bibnamefont {Liu}}, \ and\
  \bibinfo {author} {\bibfnamefont {F.}~\bibnamefont {Nori}},\ }\bibfield
  {title} {\emph {\bibinfo {title} {Generation and control of
  greenberger-horne-zeilinger entanglement in superconducting circuits},\
  }}\href {\doibase 10.1103/PhysRevLett.96.246803} {\bibfield  {journal}
  {\bibinfo  {journal} {Phys. Rev. Lett.}\ }\textbf {\bibinfo {volume} {96}},\
  \bibinfo {pages} {246803} (\bibinfo {year} {2006})}\BibitemShut {NoStop}%
\bibitem [{\citenamefont {Pan}\ \emph {et~al.}(2012{\natexlab{a}})\citenamefont
  {Pan}, \citenamefont {Chen}, \citenamefont {Lu}, \citenamefont {Weinfurter},
  \citenamefont {Zeilinger},\ and\ \citenamefont {\ifmmode~\dot{Z}\else
  \.{Z}\fi{}ukowski}}]{mp}%
  \BibitemOpen
  \bibfield  {author} {\bibinfo {author} {\bibfnamefont {J.-W.}\ \bibnamefont
  {Pan}}, \bibinfo {author} {\bibfnamefont {Z.-B.}\ \bibnamefont {Chen}},
  \bibinfo {author} {\bibfnamefont {C.-Y.}\ \bibnamefont {Lu}}, \bibinfo
  {author} {\bibfnamefont {H.}~\bibnamefont {Weinfurter}}, \bibinfo {author}
  {\bibfnamefont {A.}~\bibnamefont {Zeilinger}}, \ and\ \bibinfo {author}
  {\bibfnamefont {M.}~\bibnamefont {\ifmmode~\dot{Z}\else \.{Z}\fi{}ukowski}},\
  }\bibfield  {title} {\emph {\bibinfo {title} {Multiphoton entanglement and
  interferometry},\ }}\href {\doibase 10.1103/RevModPhys.84.777} {\bibfield
  {journal} {\bibinfo  {journal} {Rev. Mod. Phys.}\ }\textbf {\bibinfo {volume}
  {84}},\ \bibinfo {pages} {777} (\bibinfo {year}
  {2012}{\natexlab{a}})}\BibitemShut {NoStop}%
\bibitem [{\citenamefont {Tashima}\ \emph {et~al.}(2016)\citenamefont
  {Tashima}, \citenamefont {Tame}, \citenamefont {\"Ozdemir}, \citenamefont
  {Nori}, \citenamefont {Koashi},\ and\ \citenamefont {Weinfurter}}]{pm}%
  \BibitemOpen
  \bibfield  {author} {\bibinfo {author} {\bibfnamefont {T.}~\bibnamefont
  {Tashima}}, \bibinfo {author} {\bibfnamefont {M.~S.}\ \bibnamefont {Tame}},
  \bibinfo {author} {\bibfnamefont {S.~K.}\ \bibnamefont {\"Ozdemir}}, \bibinfo
  {author} {\bibfnamefont {F.}~\bibnamefont {Nori}}, \bibinfo {author}
  {\bibfnamefont {M.}~\bibnamefont {Koashi}}, \ and\ \bibinfo {author}
  {\bibfnamefont {H.}~\bibnamefont {Weinfurter}},\ }\bibfield  {title} {\emph
  {\bibinfo {title} {Photonic multipartite entanglement conversion using
  nonlocal operations},\ }}\href {\doibase 10.1103/PhysRevA.94.052309}
  {\bibfield  {journal} {\bibinfo  {journal} {Phys. Rev. A}\ }\textbf {\bibinfo
  {volume} {94}},\ \bibinfo {pages} {052309} (\bibinfo {year}
  {2016})}\BibitemShut {NoStop}%
\bibitem [{\citenamefont {Macr\`{\i}}\ \emph {et~al.}(2018)\citenamefont
  {Macr\`{\i}}, \citenamefont {Nori},\ and\ \citenamefont {Kockum}}]{bg}%
  \BibitemOpen
  \bibfield  {author} {\bibinfo {author} {\bibfnamefont {V.}~\bibnamefont
  {Macr\`{\i}}}, \bibinfo {author} {\bibfnamefont {F.}~\bibnamefont {Nori}}, \
  and\ \bibinfo {author} {\bibfnamefont {A.~F.}\ \bibnamefont {Kockum}},\
  }\bibfield  {title} {\emph {\bibinfo {title} {Simple preparation of bell and
  greenberger-horne-zeilinger states using ultrastrong-coupling circuit qed},\
  }}\href {\doibase 10.1103/PhysRevA.98.062327} {\bibfield  {journal} {\bibinfo
   {journal} {Phys. Rev. A}\ }\textbf {\bibinfo {volume} {98}},\ \bibinfo
  {pages} {062327} (\bibinfo {year} {2018})}\BibitemShut {NoStop}%
\bibitem [{\citenamefont {Boto}\ \emph {et~al.}(2000)\citenamefont {Boto},
  \citenamefont {Kok}, \citenamefont {Abrams}, \citenamefont {Braunstein},
  \citenamefont {Williams},\ and\ \citenamefont {Dowling}}]{ql1}%
  \BibitemOpen
  \bibfield  {author} {\bibinfo {author} {\bibfnamefont {A.~N.}\ \bibnamefont
  {Boto}}, \bibinfo {author} {\bibfnamefont {P.}~\bibnamefont {Kok}}, \bibinfo
  {author} {\bibfnamefont {D.~S.}\ \bibnamefont {Abrams}}, \bibinfo {author}
  {\bibfnamefont {S.~L.}\ \bibnamefont {Braunstein}}, \bibinfo {author}
  {\bibfnamefont {C.~P.}\ \bibnamefont {Williams}}, \ and\ \bibinfo {author}
  {\bibfnamefont {J.~P.}\ \bibnamefont {Dowling}},\ }\bibfield  {title} {\emph
  {\bibinfo {title} {Quantum interferometric optical lithography: Exploiting
  entanglement to beat the diffraction limit},\ }}\href {\doibase
  10.1103/PhysRevLett.85.2733} {\bibfield  {journal} {\bibinfo  {journal}
  {Phys. Rev. Lett.}\ }\textbf {\bibinfo {volume} {85}},\ \bibinfo {pages}
  {2733} (\bibinfo {year} {2000})}\BibitemShut {NoStop}%
\bibitem [{\citenamefont {D'Angelo}\ \emph {et~al.}(2001)\citenamefont
  {D'Angelo}, \citenamefont {Chekhova},\ and\ \citenamefont {Shih}}]{ql2}%
  \BibitemOpen
  \bibfield  {author} {\bibinfo {author} {\bibfnamefont {M.}~\bibnamefont
  {D'Angelo}}, \bibinfo {author} {\bibfnamefont {M.~V.}\ \bibnamefont
  {Chekhova}}, \ and\ \bibinfo {author} {\bibfnamefont {Y.}~\bibnamefont
  {Shih}},\ }\bibfield  {title} {\emph {\bibinfo {title} {Two-photon
  diffraction and quantum lithography},\ }}\href {\doibase
  10.1103/PhysRevLett.87.013602} {\bibfield  {journal} {\bibinfo  {journal}
  {Phys. Rev. Lett.}\ }\textbf {\bibinfo {volume} {87}},\ \bibinfo {pages}
  {013602} (\bibinfo {year} {2001})}\BibitemShut {NoStop}%
\bibitem [{\citenamefont {Kok}\ \emph {et~al.}(2002)\citenamefont {Kok},
  \citenamefont {Lee},\ and\ \citenamefont {Dowling}}]{qm1}%
  \BibitemOpen
  \bibfield  {author} {\bibinfo {author} {\bibfnamefont {P.}~\bibnamefont
  {Kok}}, \bibinfo {author} {\bibfnamefont {H.}~\bibnamefont {Lee}}, \ and\
  \bibinfo {author} {\bibfnamefont {J.~P.}\ \bibnamefont {Dowling}},\
  }\bibfield  {title} {\emph {\bibinfo {title} {Creation of large-photon-number
  path entanglement conditioned on photodetection},\ }}\href {\doibase
  10.1103/PhysRevA.65.052104} {\bibfield  {journal} {\bibinfo  {journal} {Phys.
  Rev. A}\ }\textbf {\bibinfo {volume} {65}},\ \bibinfo {pages} {052104}
  (\bibinfo {year} {2002})}\BibitemShut {NoStop}%
\bibitem [{\citenamefont {Mitchell}\ \emph {et~al.}(2004)\citenamefont
  {Mitchell}, \citenamefont {Lundeen},\ and\ \citenamefont {Steinberg}}]{qm2}%
  \BibitemOpen
  \bibfield  {author} {\bibinfo {author} {\bibfnamefont {M.~W.}\ \bibnamefont
  {Mitchell}}, \bibinfo {author} {\bibfnamefont {J.~S.}\ \bibnamefont
  {Lundeen}}, \ and\ \bibinfo {author} {\bibfnamefont {A.~M.}\ \bibnamefont
  {Steinberg}},\ }\bibfield  {title} {\emph {\bibinfo {title} {Super-resolving
  phase measurements with a muitiphoton entangled state},\ }}\href@noop {}
  {\bibfield  {journal} {\bibinfo  {journal} {Natrue}\ }\textbf {\bibinfo
  {volume} {429}},\ \bibinfo {pages} {161} (\bibinfo {year}
  {2004})}\BibitemShut {NoStop}%
\bibitem [{\citenamefont {Bennett}\ and\ \citenamefont
  {DiVincenzo}(2000)}]{qip}%
  \BibitemOpen
  \bibfield  {author} {\bibinfo {author} {\bibfnamefont {C.~H.}\ \bibnamefont
  {Bennett}}\ and\ \bibinfo {author} {\bibfnamefont {B.~D.}\ \bibnamefont
  {DiVincenzo}},\ }\bibfield  {title} {\emph {\bibinfo {title} {Quantum
  information and computation},\ }}\href@noop {} {\bibfield  {journal}
  {\bibinfo  {journal} {Natrue}\ }\textbf {\bibinfo {volume} {404}},\ \bibinfo
  {pages} {247} (\bibinfo {year} {2000})}\BibitemShut {NoStop}%
\bibitem [{\citenamefont {Strauch}\ \emph {et~al.}(2010)\citenamefont
  {Strauch}, \citenamefont {Jacobs},\ and\ \citenamefont {Simmonds}}]{noon1}%
  \BibitemOpen
  \bibfield  {author} {\bibinfo {author} {\bibfnamefont {F.~W.}\ \bibnamefont
  {Strauch}}, \bibinfo {author} {\bibfnamefont {K.}~\bibnamefont {Jacobs}}, \
  and\ \bibinfo {author} {\bibfnamefont {R.~W.}\ \bibnamefont {Simmonds}},\
  }\bibfield  {title} {\emph {\bibinfo {title} {Arbitrary control of
  entanglement between two superconducting resonators},\ }}\href {\doibase
  10.1103/PhysRevLett.105.050501} {\bibfield  {journal} {\bibinfo  {journal}
  {Phys. Rev. Lett.}\ }\textbf {\bibinfo {volume} {105}},\ \bibinfo {pages}
  {050501} (\bibinfo {year} {2010})}\BibitemShut {NoStop}%
\bibitem [{\citenamefont {Merkel}\ and\ \citenamefont {Wilhelm}(2010)}]{noon2}%
  \BibitemOpen
  \bibfield  {author} {\bibinfo {author} {\bibfnamefont {S.~T.}\ \bibnamefont
  {Merkel}}\ and\ \bibinfo {author} {\bibfnamefont {F.~K.}\ \bibnamefont
  {Wilhelm}},\ }\bibfield  {title} {\emph {\bibinfo {title} {Generation and
  detection of noon states in superconducting circuits},\ }}\href@noop {}
  {\bibfield  {journal} {\bibinfo  {journal} {New J. Phys.}\ }\textbf {\bibinfo
  {volume} {12}},\ \bibinfo {pages} {3175} (\bibinfo {year}
  {2010})}\BibitemShut {NoStop}%
\bibitem [{\citenamefont {Wang}\ \emph {et~al.}(2011)\citenamefont {Wang},
  \citenamefont {Mariantoni}, \citenamefont {Bialczak}, \citenamefont
  {Lenander}, \citenamefont {Lucero}, \citenamefont {Neeley}, \citenamefont
  {O'Connell}, \citenamefont {Sank}, \citenamefont {Weides}, \citenamefont
  {Wenner}, \citenamefont {Yamamoto}, \citenamefont {Yin}, \citenamefont
  {Zhao}, \citenamefont {Martinis},\ and\ \citenamefont {Cleland}}]{noon3}%
  \BibitemOpen
  \bibfield  {author} {\bibinfo {author} {\bibfnamefont {H.}~\bibnamefont
  {Wang}}, \bibinfo {author} {\bibfnamefont {M.}~\bibnamefont {Mariantoni}},
  \bibinfo {author} {\bibfnamefont {R.~C.}\ \bibnamefont {Bialczak}}, \bibinfo
  {author} {\bibfnamefont {M.}~\bibnamefont {Lenander}}, \bibinfo {author}
  {\bibfnamefont {E.}~\bibnamefont {Lucero}}, \bibinfo {author} {\bibfnamefont
  {M.}~\bibnamefont {Neeley}}, \bibinfo {author} {\bibfnamefont {A.~D.}\
  \bibnamefont {O'Connell}}, \bibinfo {author} {\bibfnamefont {D.}~\bibnamefont
  {Sank}}, \bibinfo {author} {\bibfnamefont {M.}~\bibnamefont {Weides}},
  \bibinfo {author} {\bibfnamefont {J.}~\bibnamefont {Wenner}}, \bibinfo
  {author} {\bibfnamefont {T.}~\bibnamefont {Yamamoto}}, \bibinfo {author}
  {\bibfnamefont {Y.}~\bibnamefont {Yin}}, \bibinfo {author} {\bibfnamefont
  {J.}~\bibnamefont {Zhao}}, \bibinfo {author} {\bibfnamefont {J.~M.}\
  \bibnamefont {Martinis}}, \ and\ \bibinfo {author} {\bibfnamefont {A.~N.}\
  \bibnamefont {Cleland}},\ }\bibfield  {title} {\emph {\bibinfo {title}
  {Deterministic entanglement of photons in two superconducting microwave
  resonators},\ }}\href {\doibase 10.1103/PhysRevLett.106.060401} {\bibfield
  {journal} {\bibinfo  {journal} {Phys. Rev. Lett.}\ }\textbf {\bibinfo
  {volume} {106}},\ \bibinfo {pages} {060401} (\bibinfo {year}
  {2011})}\BibitemShut {NoStop}%
\bibitem [{\citenamefont {Su}\ \emph {et~al.}(2014)\citenamefont {Su},
  \citenamefont {Yang},\ and\ \citenamefont {Zheng}}]{noon4}%
  \BibitemOpen
  \bibfield  {author} {\bibinfo {author} {\bibfnamefont {Q.~P.}\ \bibnamefont
  {Su}}, \bibinfo {author} {\bibfnamefont {C.~P.}\ \bibnamefont {Yang}}, \ and\
  \bibinfo {author} {\bibfnamefont {S.~B.}\ \bibnamefont {Zheng}},\ }\bibfield
  {title} {\emph {\bibinfo {title} {Fast and simple scheme for generating noon
  states of photons in circuit qed},\ }}\href@noop {} {\bibfield  {journal}
  {\bibinfo  {journal} {Sci. Rep.}\ }\textbf {\bibinfo {volume} {4}},\ \bibinfo
  {pages} {3898} (\bibinfo {year} {2014})}\BibitemShut {NoStop}%
\bibitem [{\citenamefont {Xiong}\ \emph {et~al.}(2015)\citenamefont {Xiong},
  \citenamefont {Sun}, \citenamefont {Liu}, \citenamefont {Liu},\ and\
  \citenamefont {Yang}}]{noon5}%
  \BibitemOpen
  \bibfield  {author} {\bibinfo {author} {\bibfnamefont {S.~J.}\ \bibnamefont
  {Xiong}}, \bibinfo {author} {\bibfnamefont {Z.}~\bibnamefont {Sun}}, \bibinfo
  {author} {\bibfnamefont {J.~M.}\ \bibnamefont {Liu}}, \bibinfo {author}
  {\bibfnamefont {T.}~\bibnamefont {Liu}}, \ and\ \bibinfo {author}
  {\bibfnamefont {C.~P.}\ \bibnamefont {Yang}},\ }\bibfield  {title} {\emph
  {\bibinfo {title} {Efficient scheme for generation of photonic noon states in
  circuit qed},\ }}\href@noop {} {\bibfield  {journal} {\bibinfo  {journal}
  {Opt. Lett.}\ }\textbf {\bibinfo {volume} {40}},\ \bibinfo {pages} {2221}
  (\bibinfo {year} {2015})}\BibitemShut {NoStop}%
\bibitem [{\citenamefont {Backens}(2017)}]{noon6}%
  \BibitemOpen
  \bibfield  {author} {\bibinfo {author} {\bibfnamefont {M.}~\bibnamefont
  {Backens}},\ }\bibfield  {title} {\emph {\bibinfo {title} {Number of
  superclasses of four-qubit entangled states under the inductive entanglement
  classification},\ }}\href {\doibase 10.1103/PhysRevA.95.022329} {\bibfield
  {journal} {\bibinfo  {journal} {Phys. Rev. A}\ }\textbf {\bibinfo {volume}
  {95}},\ \bibinfo {pages} {022329} (\bibinfo {year} {2017})}\BibitemShut
  {NoStop}%
\bibitem [{\citenamefont {Casanova}\ \emph {et~al.}(2010)\citenamefont
  {Casanova}, \citenamefont {Romero}, \citenamefont {Lizuain}, \citenamefont
  {Garc\'{\i}a-Ripoll},\ and\ \citenamefont {Solano}}]{ip}%
  \BibitemOpen
  \bibfield  {author} {\bibinfo {author} {\bibfnamefont {J.}~\bibnamefont
  {Casanova}}, \bibinfo {author} {\bibfnamefont {G.}~\bibnamefont {Romero}},
  \bibinfo {author} {\bibfnamefont {I.}~\bibnamefont {Lizuain}}, \bibinfo
  {author} {\bibfnamefont {J.~J.}\ \bibnamefont {Garc\'{\i}a-Ripoll}}, \ and\
  \bibinfo {author} {\bibfnamefont {E.}~\bibnamefont {Solano}},\ }\bibfield
  {title} {\emph {\bibinfo {title} {Deep strong coupling regime of the
  jaynes-cummings model},\ }}\href {\doibase 10.1103/PhysRevLett.105.263603}
  {\bibfield  {journal} {\bibinfo  {journal} {Phys. Rev. Lett.}\ }\textbf
  {\bibinfo {volume} {105}},\ \bibinfo {pages} {263603} (\bibinfo {year}
  {2010})}\BibitemShut {NoStop}%
\bibitem [{\citenamefont {Ai}\ \emph {et~al.}(2010)\citenamefont {Ai},
  \citenamefont {Li}, \citenamefont {Zheng},\ and\ \citenamefont {Sun}}]{ip2}%
  \BibitemOpen
  \bibfield  {author} {\bibinfo {author} {\bibfnamefont {Q.}~\bibnamefont
  {Ai}}, \bibinfo {author} {\bibfnamefont {Y.}~\bibnamefont {Li}}, \bibinfo
  {author} {\bibfnamefont {H.}~\bibnamefont {Zheng}}, \ and\ \bibinfo {author}
  {\bibfnamefont {C.~P.}\ \bibnamefont {Sun}},\ }\bibfield  {title} {\emph
  {\bibinfo {title} {Quantum anti-zeno effect without rotating wave
  approximation},\ }}\href {\doibase 10.1103/PhysRevA.81.042116} {\bibfield
  {journal} {\bibinfo  {journal} {Phys. Rev. A}\ }\textbf {\bibinfo {volume}
  {81}},\ \bibinfo {pages} {042116} (\bibinfo {year} {2010})}\BibitemShut
  {NoStop}%
\bibitem [{\citenamefont {Forn-D\'{\i}az}\ \emph {et~al.}(2010)\citenamefont
  {Forn-D\'{\i}az}, \citenamefont {Lisenfeld}, \citenamefont {Marcos},
  \citenamefont {Garc\'{\i}a-Ripoll}, \citenamefont {Solano}, \citenamefont
  {Harmans},\ and\ \citenamefont {Mooij}}]{ip3}%
  \BibitemOpen
  \bibfield  {author} {\bibinfo {author} {\bibfnamefont {P.}~\bibnamefont
  {Forn-D\'{\i}az}}, \bibinfo {author} {\bibfnamefont {J.}~\bibnamefont
  {Lisenfeld}}, \bibinfo {author} {\bibfnamefont {D.}~\bibnamefont {Marcos}},
  \bibinfo {author} {\bibfnamefont {J.~J.}\ \bibnamefont {Garc\'{\i}a-Ripoll}},
  \bibinfo {author} {\bibfnamefont {E.}~\bibnamefont {Solano}}, \bibinfo
  {author} {\bibfnamefont {C.~J. P.~M.}\ \bibnamefont {Harmans}}, \ and\
  \bibinfo {author} {\bibfnamefont {J.~E.}\ \bibnamefont {Mooij}},\ }\bibfield
  {title} {\emph {\bibinfo {title} {Observation of the bloch-siegert shift in a
  qubit-oscillator system in the ultrastrong coupling regime},\ }}\href
  {\doibase 10.1103/PhysRevLett.105.237001} {\bibfield  {journal} {\bibinfo
  {journal} {Phys. Rev. Lett.}\ }\textbf {\bibinfo {volume} {105}},\ \bibinfo
  {pages} {237001} (\bibinfo {year} {2010})}\BibitemShut {NoStop}%
\bibitem [{\citenamefont {Braak}(2011)}]{ip4}%
  \BibitemOpen
  \bibfield  {author} {\bibinfo {author} {\bibfnamefont {D.}~\bibnamefont
  {Braak}},\ }\bibfield  {title} {\emph {\bibinfo {title} {Integrability of the
  rabi model},\ }}\href {\doibase 10.1103/PhysRevLett.107.100401} {\bibfield
  {journal} {\bibinfo  {journal} {Phys. Rev. Lett.}\ }\textbf {\bibinfo
  {volume} {107}},\ \bibinfo {pages} {100401} (\bibinfo {year}
  {2011})}\BibitemShut {NoStop}%
\bibitem [{\citenamefont {Ridolfo}\ \emph {et~al.}(2013)\citenamefont
  {Ridolfo}, \citenamefont {Savasta},\ and\ \citenamefont {Hartmann}}]{ip5}%
  \BibitemOpen
  \bibfield  {author} {\bibinfo {author} {\bibfnamefont {A.}~\bibnamefont
  {Ridolfo}}, \bibinfo {author} {\bibfnamefont {S.}~\bibnamefont {Savasta}}, \
  and\ \bibinfo {author} {\bibfnamefont {M.~J.}\ \bibnamefont {Hartmann}},\
  }\bibfield  {title} {\emph {\bibinfo {title} {Nonclassical radiation from
  thermal cavities in the ultrastrong coupling regime},\ }}\href {\doibase
  10.1103/PhysRevLett.110.163601} {\bibfield  {journal} {\bibinfo  {journal}
  {Phys. Rev. Lett.}\ }\textbf {\bibinfo {volume} {110}},\ \bibinfo {pages}
  {163601} (\bibinfo {year} {2013})}\BibitemShut {NoStop}%
\bibitem [{\citenamefont {Zhao}\ \emph {et~al.}(2015)\citenamefont {Zhao},
  \citenamefont {Liu}, \citenamefont {Liu},\ and\ \citenamefont {Nori}}]{ip6}%
  \BibitemOpen
  \bibfield  {author} {\bibinfo {author} {\bibfnamefont {Y.-J.}\ \bibnamefont
  {Zhao}}, \bibinfo {author} {\bibfnamefont {Y.-L.}\ \bibnamefont {Liu}},
  \bibinfo {author} {\bibfnamefont {Y.-x.}\ \bibnamefont {Liu}}, \ and\
  \bibinfo {author} {\bibfnamefont {F.}~\bibnamefont {Nori}},\ }\bibfield
  {title} {\emph {\bibinfo {title} {Generating nonclassical photon states via
  longitudinal couplings between superconducting qubits and microwave fields},\
  }}\href {\doibase 10.1103/PhysRevA.91.053820} {\bibfield  {journal} {\bibinfo
   {journal} {Phys. Rev. A}\ }\textbf {\bibinfo {volume} {91}},\ \bibinfo
  {pages} {053820} (\bibinfo {year} {2015})}\BibitemShut {NoStop}%
\bibitem [{\citenamefont {Ashhab}(2013)}]{ip7}%
  \BibitemOpen
  \bibfield  {author} {\bibinfo {author} {\bibfnamefont {S.}~\bibnamefont
  {Ashhab}},\ }\bibfield  {title} {\emph {\bibinfo {title} {Superradiance
  transition in a system with a single qubit and a single oscillator},\ }}\href
  {\doibase 10.1103/PhysRevA.87.013826} {\bibfield  {journal} {\bibinfo
  {journal} {Phys. Rev. A}\ }\textbf {\bibinfo {volume} {87}},\ \bibinfo
  {pages} {013826} (\bibinfo {year} {2013})}\BibitemShut {NoStop}%
\bibitem [{\citenamefont {Garziano}\ \emph {et~al.}(2016)\citenamefont
  {Garziano}, \citenamefont {Macr\`{\i}}, \citenamefont {Stassi}, \citenamefont
  {Di~Stefano}, \citenamefont {Nori},\ and\ \citenamefont {Savasta}}]{ip8}%
  \BibitemOpen
  \bibfield  {author} {\bibinfo {author} {\bibfnamefont {L.}~\bibnamefont
  {Garziano}}, \bibinfo {author} {\bibfnamefont {V.}~\bibnamefont
  {Macr\`{\i}}}, \bibinfo {author} {\bibfnamefont {R.}~\bibnamefont {Stassi}},
  \bibinfo {author} {\bibfnamefont {O.}~\bibnamefont {Di~Stefano}}, \bibinfo
  {author} {\bibfnamefont {F.}~\bibnamefont {Nori}}, \ and\ \bibinfo {author}
  {\bibfnamefont {S.}~\bibnamefont {Savasta}},\ }\bibfield  {title} {\emph
  {\bibinfo {title} {One photon can simultaneously excite two or more atoms},\
  }}\href {\doibase 10.1103/PhysRevLett.117.043601} {\bibfield  {journal}
  {\bibinfo  {journal} {Phys. Rev. Lett.}\ }\textbf {\bibinfo {volume} {117}},\
  \bibinfo {pages} {043601} (\bibinfo {year} {2016})}\BibitemShut {NoStop}%
\bibitem [{\citenamefont {Pan}\ \emph {et~al.}(2012{\natexlab{b}})\citenamefont
  {Pan}, \citenamefont {Chen}, \citenamefont {Lu}, \citenamefont {Weinfurter},
  \citenamefont {Zeilinger},\ and\ \citenamefont {\ifmmode~\dot{Z}\else
  \.{Z}\fi{}ukowski}}]{ip9}%
  \BibitemOpen
  \bibfield  {author} {\bibinfo {author} {\bibfnamefont {J.-W.}\ \bibnamefont
  {Pan}}, \bibinfo {author} {\bibfnamefont {Z.-B.}\ \bibnamefont {Chen}},
  \bibinfo {author} {\bibfnamefont {C.-Y.}\ \bibnamefont {Lu}}, \bibinfo
  {author} {\bibfnamefont {H.}~\bibnamefont {Weinfurter}}, \bibinfo {author}
  {\bibfnamefont {A.}~\bibnamefont {Zeilinger}}, \ and\ \bibinfo {author}
  {\bibfnamefont {M.}~\bibnamefont {\ifmmode~\dot{Z}\else \.{Z}\fi{}ukowski}},\
  }\bibfield  {title} {\emph {\bibinfo {title} {Multiphoton entanglement and
  interferometry},\ }}\href {\doibase 10.1103/RevModPhys.84.777} {\bibfield
  {journal} {\bibinfo  {journal} {Rev. Mod. Phys.}\ }\textbf {\bibinfo {volume}
  {84}},\ \bibinfo {pages} {777} (\bibinfo {year}
  {2012}{\natexlab{b}})}\BibitemShut {NoStop}%
\bibitem [{\citenamefont {Zhao}\ \emph {et~al.}(2017)\citenamefont {Zhao},
  \citenamefont {Tan}, \citenamefont {Yu}, \citenamefont {Zhu},\ and\
  \citenamefont {Yu}}]{ip10}%
  \BibitemOpen
  \bibfield  {author} {\bibinfo {author} {\bibfnamefont {P.}~\bibnamefont
  {Zhao}}, \bibinfo {author} {\bibfnamefont {X.}~\bibnamefont {Tan}}, \bibinfo
  {author} {\bibfnamefont {H.}~\bibnamefont {Yu}}, \bibinfo {author}
  {\bibfnamefont {S.-L.}\ \bibnamefont {Zhu}}, \ and\ \bibinfo {author}
  {\bibfnamefont {Y.}~\bibnamefont {Yu}},\ }\bibfield  {title} {\emph {\bibinfo
  {title} {Circuit qed with qutrits: Coupling three or more atoms via
  virtual-photon exchange},\ }}\href {\doibase 10.1103/PhysRevA.96.043833}
  {\bibfield  {journal} {\bibinfo  {journal} {Phys. Rev. A}\ }\textbf {\bibinfo
  {volume} {96}},\ \bibinfo {pages} {043833} (\bibinfo {year}
  {2017})}\BibitemShut {NoStop}%
\bibitem [{\citenamefont {Ma}\ and\ \citenamefont {Law}(2015)}]{sc}%
  \BibitemOpen
  \bibfield  {author} {\bibinfo {author} {\bibfnamefont {K.~K.~W.}\
  \bibnamefont {Ma}}\ and\ \bibinfo {author} {\bibfnamefont {C.~K.}\
  \bibnamefont {Law}},\ }\bibfield  {title} {\emph {\bibinfo {title}
  {Three-photon resonance and adiabatic passage in the large-detuning rabi
  model},\ }}\href {\doibase 10.1103/PhysRevA.92.023842} {\bibfield  {journal}
  {\bibinfo  {journal} {Phys. Rev. A}\ }\textbf {\bibinfo {volume} {92}},\
  \bibinfo {pages} {023842} (\bibinfo {year} {2015})}\BibitemShut {NoStop}%
\bibitem [{\citenamefont {Garziano}\ \emph {et~al.}(2015)\citenamefont
  {Garziano}, \citenamefont {Stassi}, \citenamefont {Macr\`{\i}}, \citenamefont
  {Kockum}, \citenamefont {Savasta},\ and\ \citenamefont {Nori}}]{sc2}%
  \BibitemOpen
  \bibfield  {author} {\bibinfo {author} {\bibfnamefont {L.}~\bibnamefont
  {Garziano}}, \bibinfo {author} {\bibfnamefont {R.}~\bibnamefont {Stassi}},
  \bibinfo {author} {\bibfnamefont {V.}~\bibnamefont {Macr\`{\i}}}, \bibinfo
  {author} {\bibfnamefont {A.~F.}\ \bibnamefont {Kockum}}, \bibinfo {author}
  {\bibfnamefont {S.}~\bibnamefont {Savasta}}, \ and\ \bibinfo {author}
  {\bibfnamefont {F.}~\bibnamefont {Nori}},\ }\bibfield  {title} {\emph
  {\bibinfo {title} {Multiphoton quantum rabi oscillations in ultrastrong
  cavity qed},\ }}\href {\doibase 10.1103/PhysRevA.92.063830} {\bibfield
  {journal} {\bibinfo  {journal} {Phys. Rev. A}\ }\textbf {\bibinfo {volume}
  {92}},\ \bibinfo {pages} {063830} (\bibinfo {year} {2015})}\BibitemShut
  {NoStop}%
\bibitem [{\citenamefont {Stassi}\ \emph {et~al.}(2017)\citenamefont {Stassi},
  \citenamefont {Macr\`{\i}}, \citenamefont {Kockum}, \citenamefont
  {Di~Stefano}, \citenamefont {Miranowicz}, \citenamefont {Savasta},\ and\
  \citenamefont {Nori}}]{ql}%
  \BibitemOpen
  \bibfield  {author} {\bibinfo {author} {\bibfnamefont {R.}~\bibnamefont
  {Stassi}}, \bibinfo {author} {\bibfnamefont {V.}~\bibnamefont {Macr\`{\i}}},
  \bibinfo {author} {\bibfnamefont {A.~F.}\ \bibnamefont {Kockum}}, \bibinfo
  {author} {\bibfnamefont {O.}~\bibnamefont {Di~Stefano}}, \bibinfo {author}
  {\bibfnamefont {A.}~\bibnamefont {Miranowicz}}, \bibinfo {author}
  {\bibfnamefont {S.}~\bibnamefont {Savasta}}, \ and\ \bibinfo {author}
  {\bibfnamefont {F.}~\bibnamefont {Nori}},\ }\bibfield  {title} {\emph
  {\bibinfo {title} {Quantum nonlinear optics without photons},\ }}\href
  {\doibase 10.1103/PhysRevA.96.023818} {\bibfield  {journal} {\bibinfo
  {journal} {Phys. Rev. A}\ }\textbf {\bibinfo {volume} {96}},\ \bibinfo
  {pages} {023818} (\bibinfo {year} {2017})}\BibitemShut {NoStop}%
\bibitem [{\citenamefont {Wallraff}\ \emph {et~al.}(2004)\citenamefont
  {Wallraff}, \citenamefont {Schuster}, \citenamefont {Blais}, \citenamefont
  {Frunzio}, \citenamefont {R-S}, \citenamefont {Majer}, \citenamefont {Kumar},
  \citenamefont {Girvin},\ and\ \citenamefont {Schoelkopf}}]{sq}%
  \BibitemOpen
  \bibfield  {author} {\bibinfo {author} {\bibfnamefont {A.}~\bibnamefont
  {Wallraff}}, \bibinfo {author} {\bibfnamefont {D.~I.}\ \bibnamefont
  {Schuster}}, \bibinfo {author} {\bibfnamefont {A.}~\bibnamefont {Blais}},
  \bibinfo {author} {\bibfnamefont {L.}~\bibnamefont {Frunzio}}, \bibinfo
  {author} {\bibfnamefont {H.}~\bibnamefont {R-S}}, \bibinfo {author}
  {\bibfnamefont {J.}~\bibnamefont {Majer}}, \bibinfo {author} {\bibfnamefont
  {S.}~\bibnamefont {Kumar}}, \bibinfo {author} {\bibfnamefont {S.~M.}\
  \bibnamefont {Girvin}}, \ and\ \bibinfo {author} {\bibfnamefont {R.~J.}\
  \bibnamefont {Schoelkopf}},\ }\bibfield  {title} {\emph {\bibinfo {title}
  {Strong coupling of a single photon to a superconducting qubit using circuit
  quantum electrodynamics},\ }}\href@noop {} {\bibfield  {journal} {\bibinfo
  {journal} {Nature}\ }\textbf {\bibinfo {volume} {431}},\ \bibinfo {pages}
  {162} (\bibinfo {year} {2004})}\BibitemShut {NoStop}%
\bibitem [{\citenamefont {Schoelkopf}\ and\ \citenamefont
  {Girvin}(2008)}]{sq2}%
  \BibitemOpen
  \bibfield  {author} {\bibinfo {author} {\bibfnamefont {R.~J.}\ \bibnamefont
  {Schoelkopf}}\ and\ \bibinfo {author} {\bibfnamefont {S.~M.}\ \bibnamefont
  {Girvin}},\ }\bibfield  {title} {\emph {\bibinfo {title} {Wiring up quantum
  systems},\ }}\href@noop {} {\bibfield  {journal} {\bibinfo  {journal}
  {Nature}\ }\textbf {\bibinfo {volume} {451}},\ \bibinfo {pages} {664}
  (\bibinfo {year} {2008})}\BibitemShut {NoStop}%
\bibitem [{\citenamefont {Peropadre}\ \emph {et~al.}(2010)\citenamefont
  {Peropadre}, \citenamefont {Forn-D\'{\i}az}, \citenamefont {Solano},\ and\
  \citenamefont {Garc\'{\i}a-Ripoll}}]{sq3}%
  \BibitemOpen
  \bibfield  {author} {\bibinfo {author} {\bibfnamefont {B.}~\bibnamefont
  {Peropadre}}, \bibinfo {author} {\bibfnamefont {P.}~\bibnamefont
  {Forn-D\'{\i}az}}, \bibinfo {author} {\bibfnamefont {E.}~\bibnamefont
  {Solano}}, \ and\ \bibinfo {author} {\bibfnamefont {J.~J.}\ \bibnamefont
  {Garc\'{\i}a-Ripoll}},\ }\bibfield  {title} {\emph {\bibinfo {title}
  {Switchable ultrastrong coupling in circuit qed},\ }}\href {\doibase
  10.1103/PhysRevLett.105.023601} {\bibfield  {journal} {\bibinfo  {journal}
  {Phys. Rev. Lett.}\ }\textbf {\bibinfo {volume} {105}},\ \bibinfo {pages}
  {023601} (\bibinfo {year} {2010})}\BibitemShut {NoStop}%
\bibitem [{\citenamefont {You}\ and\ \citenamefont {Franco}(2011)}]{sq4}%
  \BibitemOpen
  \bibfield  {author} {\bibinfo {author} {\bibfnamefont {J.~Q.}\ \bibnamefont
  {You}}\ and\ \bibinfo {author} {\bibfnamefont {N.}~\bibnamefont {Franco}},\
  }\bibfield  {title} {\emph {\bibinfo {title} {Atomic physics and quantum
  optics using superconducting circuits},\ }}\href@noop {} {\bibfield
  {journal} {\bibinfo  {journal} {Nature}\ }\textbf {\bibinfo {volume} {474}},\
  \bibinfo {pages} {589} (\bibinfo {year} {2011})}\BibitemShut {NoStop}%
\bibitem [{\citenamefont {Niemczyk}\ \emph {et~al.}(2010)\citenamefont
  {Niemczyk}, \citenamefont {Deppe}, \citenamefont {Huebl}, \citenamefont
  {Menzel}, \citenamefont {Hocke}, \citenamefont {Schwarz}, \citenamefont
  {Garciaripoll}, \citenamefont {Zueco}, \citenamefont {H\"ommer},\ and\
  \citenamefont {Solano}}]{cq}%
  \BibitemOpen
  \bibfield  {author} {\bibinfo {author} {\bibfnamefont {T.}~\bibnamefont
  {Niemczyk}}, \bibinfo {author} {\bibfnamefont {F.}~\bibnamefont {Deppe}},
  \bibinfo {author} {\bibfnamefont {H.}~\bibnamefont {Huebl}}, \bibinfo
  {author} {\bibfnamefont {E.~P.}\ \bibnamefont {Menzel}}, \bibinfo {author}
  {\bibfnamefont {F.}~\bibnamefont {Hocke}}, \bibinfo {author} {\bibfnamefont
  {M.~J.}\ \bibnamefont {Schwarz}}, \bibinfo {author} {\bibfnamefont {J.~J.}\
  \bibnamefont {Garciaripoll}}, \bibinfo {author} {\bibfnamefont
  {D.}~\bibnamefont {Zueco}}, \bibinfo {author} {\bibfnamefont
  {T.}~\bibnamefont {H\"ommer}}, \ and\ \bibinfo {author} {\bibfnamefont
  {E.}~\bibnamefont {Solano}},\ }\bibfield  {title} {\emph {\bibinfo {title}
  {Circuit quantum electrodynamics in the ultrastrong-coupling regime},\
  }}\href@noop {} {\bibfield  {journal} {\bibinfo  {journal} {Nat. Phys.}\
  }\textbf {\bibinfo {volume} {6}},\ \bibinfo {pages} {772} (\bibinfo {year}
  {2010})}\BibitemShut {NoStop}%
\bibitem [{\citenamefont {Beaudoin}\ \emph {et~al.}(2011)\citenamefont
  {Beaudoin}, \citenamefont {Gambetta},\ and\ \citenamefont {Blais}}]{me1}%
  \BibitemOpen
  \bibfield  {author} {\bibinfo {author} {\bibfnamefont {F.}~\bibnamefont
  {Beaudoin}}, \bibinfo {author} {\bibfnamefont {J.~M.}\ \bibnamefont
  {Gambetta}}, \ and\ \bibinfo {author} {\bibfnamefont {A.}~\bibnamefont
  {Blais}},\ }\bibfield  {title} {\emph {\bibinfo {title} {Dissipation and
  ultrastrong coupling in circuit qed},\ }}\href {\doibase
  10.1103/PhysRevA.84.043832} {\bibfield  {journal} {\bibinfo  {journal} {Phys.
  Rev. A}\ }\textbf {\bibinfo {volume} {84}},\ \bibinfo {pages} {043832}
  (\bibinfo {year} {2011})}\BibitemShut {NoStop}%
\bibitem [{\citenamefont {Ridolfo}\ \emph {et~al.}(2012)\citenamefont
  {Ridolfo}, \citenamefont {Leib}, \citenamefont {Savasta},\ and\ \citenamefont
  {Hartmann}}]{me2}%
  \BibitemOpen
  \bibfield  {author} {\bibinfo {author} {\bibfnamefont {A.}~\bibnamefont
  {Ridolfo}}, \bibinfo {author} {\bibfnamefont {M.}~\bibnamefont {Leib}},
  \bibinfo {author} {\bibfnamefont {S.}~\bibnamefont {Savasta}}, \ and\
  \bibinfo {author} {\bibfnamefont {M.~J.}\ \bibnamefont {Hartmann}},\
  }\bibfield  {title} {\emph {\bibinfo {title} {Photon blockade in the
  ultrastrong coupling regime},\ }}\href {\doibase
  10.1103/PhysRevLett.109.193602} {\bibfield  {journal} {\bibinfo  {journal}
  {Phys. Rev. Lett.}\ }\textbf {\bibinfo {volume} {109}},\ \bibinfo {pages}
  {193602} (\bibinfo {year} {2012})}\BibitemShut {NoStop}%
\bibitem [{\citenamefont {Yan}\ \emph {et~al.}(2015)\citenamefont {Yan},
  \citenamefont {Gustavsson}, \citenamefont {Kamal}, \citenamefont {Birenbaum},
  \citenamefont {Sears}, \citenamefont {Hover}, \citenamefont {Gudmundsen},
  \citenamefont {Rosenberg}, \citenamefont {Samach},\ and\ \citenamefont
  {Weber}}]{fq}%
  \BibitemOpen
  \bibfield  {author} {\bibinfo {author} {\bibfnamefont {F.}~\bibnamefont
  {Yan}}, \bibinfo {author} {\bibfnamefont {S.}~\bibnamefont {Gustavsson}},
  \bibinfo {author} {\bibfnamefont {A.}~\bibnamefont {Kamal}}, \bibinfo
  {author} {\bibfnamefont {J.}~\bibnamefont {Birenbaum}}, \bibinfo {author}
  {\bibfnamefont {A.~P.}\ \bibnamefont {Sears}}, \bibinfo {author}
  {\bibfnamefont {D.}~\bibnamefont {Hover}}, \bibinfo {author} {\bibfnamefont
  {T.~J.}\ \bibnamefont {Gudmundsen}}, \bibinfo {author} {\bibfnamefont
  {D.}~\bibnamefont {Rosenberg}}, \bibinfo {author} {\bibfnamefont
  {G.}~\bibnamefont {Samach}}, \ and\ \bibinfo {author} {\bibfnamefont
  {S.}~\bibnamefont {Weber}},\ }\bibfield  {title} {\emph {\bibinfo {title}
  {The flux qubit revisited to enhance coherence and reproducibility},\
  }}\href@noop {} {\bibfield  {journal} {\bibinfo  {journal} {Nat. Commun.}\
  }\textbf {\bibinfo {volume} {7}},\ \bibinfo {pages} {12964} (\bibinfo {year}
  {2015})}\BibitemShut {NoStop}%
\bibitem [{\citenamefont {You}\ \emph {et~al.}(2007)\citenamefont {You},
  \citenamefont {Hu}, \citenamefont {Ashhab},\ and\ \citenamefont
  {Nori}}]{fq2}%
  \BibitemOpen
  \bibfield  {author} {\bibinfo {author} {\bibfnamefont {J.~Q.}\ \bibnamefont
  {You}}, \bibinfo {author} {\bibfnamefont {X.}~\bibnamefont {Hu}}, \bibinfo
  {author} {\bibfnamefont {S.}~\bibnamefont {Ashhab}}, \ and\ \bibinfo {author}
  {\bibfnamefont {F.}~\bibnamefont {Nori}},\ }\bibfield  {title} {\emph
  {\bibinfo {title} {Low-decoherence flux qubit},\ }}\href {\doibase
  10.1103/PhysRevB.75.140515} {\bibfield  {journal} {\bibinfo  {journal} {Phys.
  Rev. B}\ }\textbf {\bibinfo {volume} {75}},\ \bibinfo {pages} {140515}
  (\bibinfo {year} {2007})}\BibitemShut {NoStop}%
\bibitem [{\citenamefont {Peterer}\ \emph {et~al.}(2015)\citenamefont
  {Peterer}, \citenamefont {Bader}, \citenamefont {Jin}, \citenamefont {Yan},
  \citenamefont {Kamal}, \citenamefont {Gudmundsen}, \citenamefont {Leek},
  \citenamefont {Orlando}, \citenamefont {Oliver},\ and\ \citenamefont
  {Gustavsson}}]{fq3}%
  \BibitemOpen
  \bibfield  {author} {\bibinfo {author} {\bibfnamefont {M.~J.}\ \bibnamefont
  {Peterer}}, \bibinfo {author} {\bibfnamefont {S.~J.}\ \bibnamefont {Bader}},
  \bibinfo {author} {\bibfnamefont {X.}~\bibnamefont {Jin}}, \bibinfo {author}
  {\bibfnamefont {F.}~\bibnamefont {Yan}}, \bibinfo {author} {\bibfnamefont
  {A.}~\bibnamefont {Kamal}}, \bibinfo {author} {\bibfnamefont {T.~J.}\
  \bibnamefont {Gudmundsen}}, \bibinfo {author} {\bibfnamefont {P.~J.}\
  \bibnamefont {Leek}}, \bibinfo {author} {\bibfnamefont {T.~P.}\ \bibnamefont
  {Orlando}}, \bibinfo {author} {\bibfnamefont {W.~D.}\ \bibnamefont {Oliver}},
  \ and\ \bibinfo {author} {\bibfnamefont {S.}~\bibnamefont {Gustavsson}},\
  }\bibfield  {title} {\emph {\bibinfo {title} {Coherence and decay of higher
  energy levels of a superconducting transmon qubit},\ }}\href {\doibase
  10.1103/PhysRevLett.114.010501} {\bibfield  {journal} {\bibinfo  {journal}
  {Phys. Rev. Lett.}\ }\textbf {\bibinfo {volume} {114}},\ \bibinfo {pages}
  {010501} (\bibinfo {year} {2015})}\BibitemShut {NoStop}%
\bibitem [{\citenamefont {Pop}\ \emph {et~al.}(2014)\citenamefont {Pop},
  \citenamefont {Kurtis}, \citenamefont {Gianluigi}, \citenamefont
  {Schoelkopf}, \citenamefont {Glazman},\ and\ \citenamefont {Devoret}}]{fq4}%
  \BibitemOpen
  \bibfield  {author} {\bibinfo {author} {\bibfnamefont {I.~M.}\ \bibnamefont
  {Pop}}, \bibinfo {author} {\bibfnamefont {G.}~\bibnamefont {Kurtis}},
  \bibinfo {author} {\bibfnamefont {C.}~\bibnamefont {Gianluigi}}, \bibinfo
  {author} {\bibfnamefont {R.~J.}\ \bibnamefont {Schoelkopf}}, \bibinfo
  {author} {\bibfnamefont {L.~I.}\ \bibnamefont {Glazman}}, \ and\ \bibinfo
  {author} {\bibfnamefont {M.~H.}\ \bibnamefont {Devoret}},\ }\bibfield
  {title} {\emph {\bibinfo {title} {Coherent suppression of electromagnetic
  dissipation due to superconducting quasiparticles},\ }}\href@noop {}
  {\bibfield  {journal} {\bibinfo  {journal} {Nature}\ }\textbf {\bibinfo
  {volume} {508}},\ \bibinfo {pages} {369} (\bibinfo {year}
  {2014})}\BibitemShut {NoStop}%
\bibitem [{\citenamefont {Peropadre}\ \emph {et~al.}(2013)\citenamefont
  {Peropadre}, \citenamefont {Zueco}, \citenamefont {Wulschner}, \citenamefont
  {Deppe}, \citenamefont {Marx}, \citenamefont {Gross},\ and\ \citenamefont
  {Garc\'{\i}a-Ripoll}}]{crosstalk}%
  \BibitemOpen
  \bibfield  {author} {\bibinfo {author} {\bibfnamefont {B.}~\bibnamefont
  {Peropadre}}, \bibinfo {author} {\bibfnamefont {D.}~\bibnamefont {Zueco}},
  \bibinfo {author} {\bibfnamefont {F.}~\bibnamefont {Wulschner}}, \bibinfo
  {author} {\bibfnamefont {F.}~\bibnamefont {Deppe}}, \bibinfo {author}
  {\bibfnamefont {A.}~\bibnamefont {Marx}}, \bibinfo {author} {\bibfnamefont
  {R.}~\bibnamefont {Gross}}, \ and\ \bibinfo {author} {\bibfnamefont {J.~J.}\
  \bibnamefont {Garc\'{\i}a-Ripoll}},\ }\bibfield  {title} {\emph {\bibinfo
  {title} {Tunable coupling engineering between superconducting resonators:
  From sidebands to effective gauge fields},\ }}\href {\doibase
  10.1103/PhysRevB.87.134504} {\bibfield  {journal} {\bibinfo  {journal} {Phys.
  Rev. B}\ }\textbf {\bibinfo {volume} {87}},\ \bibinfo {pages} {134504}
  (\bibinfo {year} {2013})}\BibitemShut {NoStop}%
\bibitem [{\citenamefont {Leek}\ \emph {et~al.}(2010)\citenamefont {Leek},
  \citenamefont {Baur}, \citenamefont {Fink}, \citenamefont {Bianchetti},
  \citenamefont {Steffen}, \citenamefont {Filipp},\ and\ \citenamefont
  {Wallraff}}]{ca}%
  \BibitemOpen
  \bibfield  {author} {\bibinfo {author} {\bibfnamefont {P.~J.}\ \bibnamefont
  {Leek}}, \bibinfo {author} {\bibfnamefont {M.}~\bibnamefont {Baur}}, \bibinfo
  {author} {\bibfnamefont {J.~M.}\ \bibnamefont {Fink}}, \bibinfo {author}
  {\bibfnamefont {R.}~\bibnamefont {Bianchetti}}, \bibinfo {author}
  {\bibfnamefont {L.}~\bibnamefont {Steffen}}, \bibinfo {author} {\bibfnamefont
  {S.}~\bibnamefont {Filipp}}, \ and\ \bibinfo {author} {\bibfnamefont
  {A.}~\bibnamefont {Wallraff}},\ }\bibfield  {title} {\emph {\bibinfo {title}
  {Cavity quantum electrodynamics with separate photon storage and qubit
  readout modes},\ }}\href {\doibase 10.1103/PhysRevLett.104.100504} {\bibfield
   {journal} {\bibinfo  {journal} {Phys. Rev. Lett.}\ }\textbf {\bibinfo
  {volume} {104}},\ \bibinfo {pages} {100504} (\bibinfo {year}
  {2010})}\BibitemShut {NoStop}%
\bibitem [{\citenamefont {Yang}\ \emph {et~al.}(2012)\citenamefont {Yang},
  \citenamefont {Su},\ and\ \citenamefont {Han}}]{gab}%
  \BibitemOpen
  \bibfield  {author} {\bibinfo {author} {\bibfnamefont {C.-P.}\ \bibnamefont
  {Yang}}, \bibinfo {author} {\bibfnamefont {Q.-P.}\ \bibnamefont {Su}}, \ and\
  \bibinfo {author} {\bibfnamefont {S.}~\bibnamefont {Han}},\ }\bibfield
  {title} {\emph {\bibinfo {title} {Generation of greenberger-horne-zeilinger
  entangled states of photons in multiple cavities via a superconducting qutrit
  or an atom through resonant interaction},\ }}\href {\doibase
  10.1103/PhysRevA.86.022329} {\bibfield  {journal} {\bibinfo  {journal} {Phys.
  Rev. A}\ }\textbf {\bibinfo {volume} {86}},\ \bibinfo {pages} {022329}
  (\bibinfo {year} {2012})}\BibitemShut {NoStop}%
\end{thebibliography}%

\end{document}